\documentclass[%
 reprint,
 superscriptaddress,
 amsmath,amssymb,
 aps,
prb,
]{revtex4-2}

\usepackage{graphicx} 
\usepackage{dcolumn} 
\usepackage{bm}
\usepackage{hyperref}
\usepackage{tikz}
\usetikzlibrary{calc}

\usepackage[export]{adjustbox}

\usepackage{ellipsis} 

\usepackage{tikz}
\usetikzlibrary{calc}
\newcolumntype{M}[1]{>{\centering\arraybackslash}m{#1}}

\usepackage[final]{changes}

\hypersetup{
    colorlinks=true,
    linkcolor=red,
    filecolor=magenta,      
    urlcolor=cyan,
}
\usepackage{braket}
\usepackage{physics}
\usepackage{chemformula}

\usepackage{bm}

\newcommand{\unlscha}{u}

\newcommand{\Ynlscha}[2]{\Upsilon_{\text{nl},#1#2}}

\newcommand{\Yinvnlscha}[2]{\overset{-1}{\Upsilon}_{\text{nl},#1#2}}
\newcommand{\Ybmnlscha}{\bm{\Upsilon}_\text{nl}}
\newcommand{\Yinvbmnlscha}{\overset{-1}{\bm{\Upsilon}}_\text{nl}}

\newcommand{\Abmnlscha}{\bm{A}_\text{nl}}

\newcommand{\Tbmnlscha}{\bm{\Theta}_\text{nl}}

\newcommand{\FCbmnlscha}{\bm{\Phi}_\text{nl}}

\newcommand{\Dbmnlscha}{\bm{D}_\text{nl}}

\newcommand{\masstns}[1]{\mathcal{M}_{#1}}
\newcommand{\masstnsbm}{\bm{\mathcal{M}}}
\newcommand{\sqrtmasstnsbm}{\sqrt{\bm{\mathcal{M}}}}
\newcommand{\sqrtmasstnsTbm}{\sqrt{\bm{\mathcal{M}}}^T}
\newcommand{\sqrtmasstns}[1]{\sqrt{\mathcal{M}}_{#1}}
\newcommand{\sqrtmasstnsT}[1]{\sqrt{\mathcal{M}}^T_{#1}}

\newcommand{\invsqrtmasstnsbm}{\overset{-1}{\sqrt{\bm{\mathcal{M}}}}}
\newcommand{\invsqrtmasstnsTbm}{\overset{-T}{\sqrt{\bm{\mathcal{M}}}}}

\newcommand{\Anlschaold}{A_\text{nl}}

\newcommand{\Tnlschaold}{\Theta_\text{nl}}

\newcommand{\Dnlschaold}{D_\text{nl}}

\newcommand{\onlschaold}{\omega_\text{nl,}}
\newcommand{\onlscha}[1]{\omega_{\text{nl,}#1}}

\newcommand{\nnlscha}[1]{n_{\text{nl,}#1}}

\newcommand{\dlogJdqbm}{\bm{d}}

\newcommand{\gnlscha}[2]{g{}^{#1#2}}

\newcommand{\gbmnlscha}{\bm{g}}

\newcommand{\polnlscha}[2]{e_{\text{nl},#1}^{#2}}
\newcommand{\polbmnlscha}[1]{\bm{e}_{\text{nl},#1}}

\newcommand{\RCnlscha}{\bm{\mathcal{R}}_{\text{C}}}
\newcommand{\RTnlscha}{\bm{\mathcal{R}}_{\text{T}}}
\newcommand{\xcnlscha}{x_\text{C}}
\newcommand{\ycnlscha}{y_\text{C}}
\newcommand{\rOnlscha}{r_0}
\newcommand{\phiOnlscha}{\phi_0}
\newcommand{\RTcmpnlscha}[1]{\mathcal{R}_{\text{T},#1}}
\newcommand{\Rallnlscha}{\bm{\Gamma}_{\text{nl}}}

\newcommand{\Jnlscha}[2]{J^{#1}_{#2}}
\newcommand{\Jbmnlscha}{\bm{J}}

\newcommand{\detJnlscha}{\mathcal{J}}

\newcommand{\xibmnlscha}{\bm{\xi}}

\newcommand{\myint}{\int_{-\infty}^{+\infty} \hspace{-0.2cm}}

\newcommand{\Fnl}{F_{\text{nl}}}

\newcommand{\Knlscha}{\mathcal{K}}

\newcommand{\gaussnlscha}{\overline{\rho}_{\text{nl}}}

\newcommand{\Hhatnlscha}{\hat{\overline{\mathcal{H}}}_\text{nl}}
\newcommand{\Znlscha}{\overline{\mathcal{Z}}_\text{nl}}
\newcommand{\rhohatnlscha}{\hat{\rho}_{\text{nl}}}

\newcommand{\Kzeronlschaold}{\mathcal{K}^{(0)} (\bm{\unlscha})}

\newcommand{\Lonenlscha}[1]{\mathcal{L}^{(1)}_{#1}}
\newcommand{\Ltwonlscha}[1]{\mathcal{L}^{(2)}_{#1}}

\newcommand{\Lonebmnlscha}{\bm{\mathcal{L}}^{(1)}}
\newcommand{\Ltwobmnlscha}{\bm{\mathcal{L}}^{(2)}}

\newcommand{\Ktwonlscha}[1]{\mathcal{K}^{(2)}_{#1} }
\newcommand{\Konenlscha}[1]{\mathcal{K}^{(1)}_{#1} }

\newcommand{\Ktwobmnlscha}{\bm{\mathcal{K}}^{(2)} }
\newcommand{\Konebmnlscha}{\bm{\mathcal{K}}^{(1)} }

\newcommand{\Kzeronlscha}{\mathcal{K}^{(0)}}

\newcommand{\hfunnlscha}{h(\bm{\unlscha})}
\newcommand{\fnlscha}{f(\unlscha_2)}
\newcommand{\fsquarenlscha}{f^2(\unlscha_2)}

\newcommand{\averagegaussnl}[1]{\left\langle #1 \right\rangle_\text{nl}}

\begin{document}

\preprint{APS/123-QED}

\title{Beyond Gaussian fluctuations of quantum anharmonic nuclei.\\ The case of rotational degrees of freedom}
\author{Antonio Siciliano}
\email[]{antonio.siciliano@uniroma1.it}
\affiliation{Dipartimento di Fisica, Università di Roma La Sapienza, Piazzale Aldo Moro 5, 00185 Roma, Italy}
\author{Lorenzo Monacelli}
\affiliation{Dipartimento di Fisica, Università di Roma La Sapienza, Piazzale Aldo Moro 5, 00185 Roma, Italy}
\affiliation{Theory and Simulation of Materials (THEOS), and National Centre for Computational Design
and Discovery of Novel Materials (MARVEL), École Polytechnique Fédérale de Lausanne, 1015
Lausanne, Switzerland}
\author{Francesco Mauri}
\affiliation{Dipartimento di Fisica, Università di Roma La Sapienza, Piazzale Aldo Moro 5, 00185 Roma, Italy}

\date{\today}

\begin{abstract}
The atomic motion in molecular crystals, such as high-pressure hydrogen or hybrid organic-inorganic perovskites, is very complex due to quantum anharmonic effects. In addition, these materials accommodate rotational degrees of freedom. All the approximate methods that describe the nuclei thermodynamics using Cartesian coordinates lead to an unphysical hybridization of roto-librations with other high-energy modes. Hence, they do not accurately account for the free energy contributions of these degrees of freedom. So, a reliable description of a molecular crystal's phase diagram is only possible with Path Integral Molecular Dynamics (PIMD) at a high computational cost. This work shows how to include roto-librational modes in the Self-Consistent Harmonic Approximation (SCHA) framework. SCHA approximates the nuclei Cartesian fluctuations to be Gaussian, thus neglecting curvilinear motion. Keeping its low computational cost, we employ the generalization of SCHA, called nonlinear SCHA (NLSCHA). Our method relies on a Gaussian \textit{ansatz} for the nuclei density matrix on a curved manifold, allowing us to map roto-librations into harmonic modes defined on a surface. By optimizing the surface's curvature variationally, we minimize the free energy, allowing the spontaneous activation of these degrees of freedom without external parameters. Notably, in the limit of vanishing curvature, we recover the standard SCHA.

\end{abstract}

\maketitle


\section{Introduction}
\label{Introduction}
Thanks to the recent methodological advantages \cite{SCHA_main,Tadano_SCF3_ISCP,TDEP,SCALID,VSCF} in the field of computational condensed matter, the pivotal role of quantum fluctuations, anharmonic effects, and finite temperature excitations on the equilibrium ionic properties has been unveiled. We emphasize that a reliable free-energy calculation should encompass all degrees of freedom. Indeed, the crystal configuration can accommodate new types of atomic motion as it changes with temperature and pressure. 

In simplest structures, the only degrees of freedom are those we refer to as 'linear vibrations', i.e.\ Gaussian fluctuations in a Cartesian space. In this case, the lattice excitations are defined in a flat space, as breathing modes or molecular stretching. Certain materials exhibit rotations and librations, i.e.\ partial rotations, showcasing unique characteristics as atom movement is confined to a curved surface. Indeed, the free rotation of a diatomic molecule is the correlated motion of two atoms on a sphere, where the diameter corresponds to the average bond length. The situation is more complex in the case of a molecular crystal, where a group of atoms forms a rigid structure strongly bonded. The latter displays low-energy modes as it can rotate freely or partially without distortions. This type of motion has a low impact on the internal energy but, on the contrary, makes a crucial contribution to the total entropy, increasing the phase space available for the system. If we miss these degrees of freedom, we may not detect phase transitions driven by internal energy and entropy competition.

A prototypical example is found in the Rigid Unit Modes (RUMs) of framework materials \cite{RUMS}, formed by stiff connected polyhedrons of atoms, e.g.\ \ch{SiO4} and \ch{AlO4} tetrahedra. RUMs have been identified as the soft modes responsible for the displacive transitions, where the rigid units rotate and translate from one phase to another \cite{RUMSphasetransition}.
Similar behavior is shown by the methyl group \ch{CH3} as it can behave as a spinning top \cite{ceriotti_anharmonic_free_energies,PRLParacetamol} depending on the environment surrounding it. This molecule is found in many pharmaceutical products, e.g.\ paracetamol, and biological compounds, such as proteins and deoxyribonucleic acid (DNA).
Another interesting case is the high-pressure hydrogen phase diagram. In phase I, the \ch{H2} centers form an h.c.p.\ lattice, and the quantum distribution of protons is almost spherical, suggesting that the molecules behave as free rotators \cite{PIMD_I_II}. The increasing pressure leads to a larger intermolecular interaction, and phase II is stabilized at around $110$ GPa \cite{ScandoloPhaseII}. Smaller molecular librations replace free rotations, hence the orientational disorder is reduced. Rotations are also thermally activated as in the III-IV phase transition where the molecules behave as free rotators above $300$ K and $220$ GPa \cite{HydrogenIVMorresi}.

If used for a systematic and unbiased investigation of molecular crystal phase diagrams, a reliable free energy method should accurately describe linear vibrations and roto-librational modes. Only Molecular Dynamics (MD) simulations can achieve this at the classical level. Indeed, MD offers the advantage of including non-perturbative anharmonic effects while accurately representing all the degrees of freedom, including rotations. Nevertheless, it does not account for quantum effects, which Path-Integral Molecular Dynamics (PIMD), the non-perturbative, reference-exact method, includes at finite temperatures. However, PIMD carries a high computational cost, requiring the simultaneous evolution of a number of the system's replicas inversely proportional to temperature. For this reason, the development of approximate ionic free energy methods became of paramount importance. 

The most commonly used and straightforward is the harmonic approximation (HA). The HA fails when phonon interactions dominate due to substantial deviations of atoms from equilibrium positions, and extensive regions of the BO energy surface (BOES) are explored. Such a condition is relevant for light nuclei, leading to sizeable zero-point motion (ZPM), near the melting point, or during a second-order phase transition. In addition, the HA is not suited for molecular crystals as it completely misses rotational modes. Indeed, in this case, the small oscillations assumption of the HA breaks down, meaning that it completely disregards the rotations of molecules or groups of atoms within a molecular crystal. 

If sufficiently small, anharmonic corrections can be incorporated via perturbation theory starting from the HA. However, this approach becomes cumbersome and is ineffective for hydride superconductors, ferroelectric, and charge-density wave compounds. The growing interest in these materials prompted the community to develop other methods, Self-Consistent Phonons (SCP) theories \cite{SCHA_main,Tadano_SCF3_ISCP,SCALID}, Vibrational Self-Consistent Field (VSCF) \cite{VSCF} and the Temperature-Dependent Effective Potential (TDEP) method \cite{TDEP}. All of them, at different levels of approximation, describe anharmonicity for linear vibrations.  However, none of these methods can effectively describe roto-librations, as they all depend on Cartesian coordinates, which are insufficient for capturing the atomic motion on a surface. So, the call for approximate methods to handle rotations is more urgent than ever.

The self-consistent description of anharmonic phonons \cite{SCHA_main,SCP1,SCPII,Tadano_nomodemixing,SCP_revisited_II,TDSCHA_monacelli,TDSCHA_mio,LihmTDSCHA} is a family of variational methods that constrain the nuclear probability distribution to be Gaussian in Cartesian coordinates. This approximation allows the enforcement of symmetries, which is an excellent help for identifying phase transition, and interpolation techniques are available to describe the thermodynamic limit, avoiding the simulation of big supercells. Remarkably, none of these advantages are present in MD/PIMD. Nevertheless, when atoms rotate freely or partially, the probability distribution deviates significantly from a Gaussian shape \cite{ceriotti_anharmonic_free_energies,HydrogenIVMorresi}. Therefore, self-consistent phonon (SCP) theories \cite{SCHA_main,SCP1,SCPII,Tadano_nomodemixing,SCP_revisited_II} are not reliable in such cases.

Here, we show how the Nonlinear Self-Consistent Harmonic Approximation (NLSCHA) \cite{INPREPARAZIONE} can overcome this problem. We employ an \textit{ad-hoc} change of variables to deform the Gaussian \textit{ansatz} and allow normal modes to occur in a curved manifold, capturing the rotations of molecules and rigid body clusters. The key aspect is that we variationally optimize the curvature, allowing the system to spontaneously activate these modes only if the quantum free energy is minimized. Remarkably, if the curvature is zero, the surface on which we constrain the atomic motion becomes flat, and in this limit, we describe only linear vibrations with the same accuracy as standard SCP theories.

In section \ref{SEC: Failure on rotational modes}, we discuss the failure of SCP theories \cite{SCHA_main,Tadano_SCF3_ISCP,SCALID} in accounting for roto-librational degrees of freedom with a 2D model for \ch{H2}, as already noted in Refs \cite{ceriotti_anharmonic_free_energies, HydrogenIVMorresi}. To address such limitation, section \ref{SEC: The solution of nonlinear Self-Consistent Harmonic Approximation} shows how to incorporate these modes in the NLSCHA framework with negligible computational cost. In the end, in sections \ref{SEC: Results at zero temperature}-\ref{SEC: Phase transition}, we benchmark our method at zero and finite temperature, and in section \ref{SECTION: 3D case} we show how to generalize our method to the three-dimensional case of molecules and crystals.

\section{Failure on rotational modes}
\label{SEC: Failure on rotational modes}
Here we show the failure of the SCP methods with molecular rotations. In particular, among these methods, we consider the Self-Consistent Harmonic Approximation (SCHA) \cite{HootonSCHA,SCHA_main}. 
To this purpose, we solve the \ch{H2} molecule rotating in two dimensions. In the center of mass reference frame, the only degree of freedom is the relative coordinate, $\bm{R} =\bm{R}_1-\bm{R}_2 = (x, y)$, with an effective mass $m=m_{\ch{H}}/2$. The BOES is given by a Morse potential fitted ab initio with DFT-BLYP on $\ch{H2}$ (see appendix \ref{APP: Exact diagonalization}) plus an empirical crystal field $E$ along the $x$ direction to control the rotational disorder
\begin{equation}
\label{def: toy model potential}
    V^{\text{(BO)}}(\bm{R}) = V_0 + d\left\{1 - \exp{-a(|\bm{R}| - R_\text{eq})}\right\}+ E x
\end{equation}
where $|\bm{R}| = \sqrt{x^2 + y^2}$
\begin{align}
\label{eq: toy model free parameters}
    V_0 =&  -1.172 \text{ Ha} &  d =  & 0.137 \text{ Ha} \\
    a = & 1.217 \text{ Bohr}^{-1} &  R_\text{eq} = & 1.393 \text{ Bohr} \notag
\end{align}

\begin{figure}[!htb]
    \centering
    \begin{minipage}[c]{1.0\linewidth}
    \includegraphics[width=1.0\textwidth]{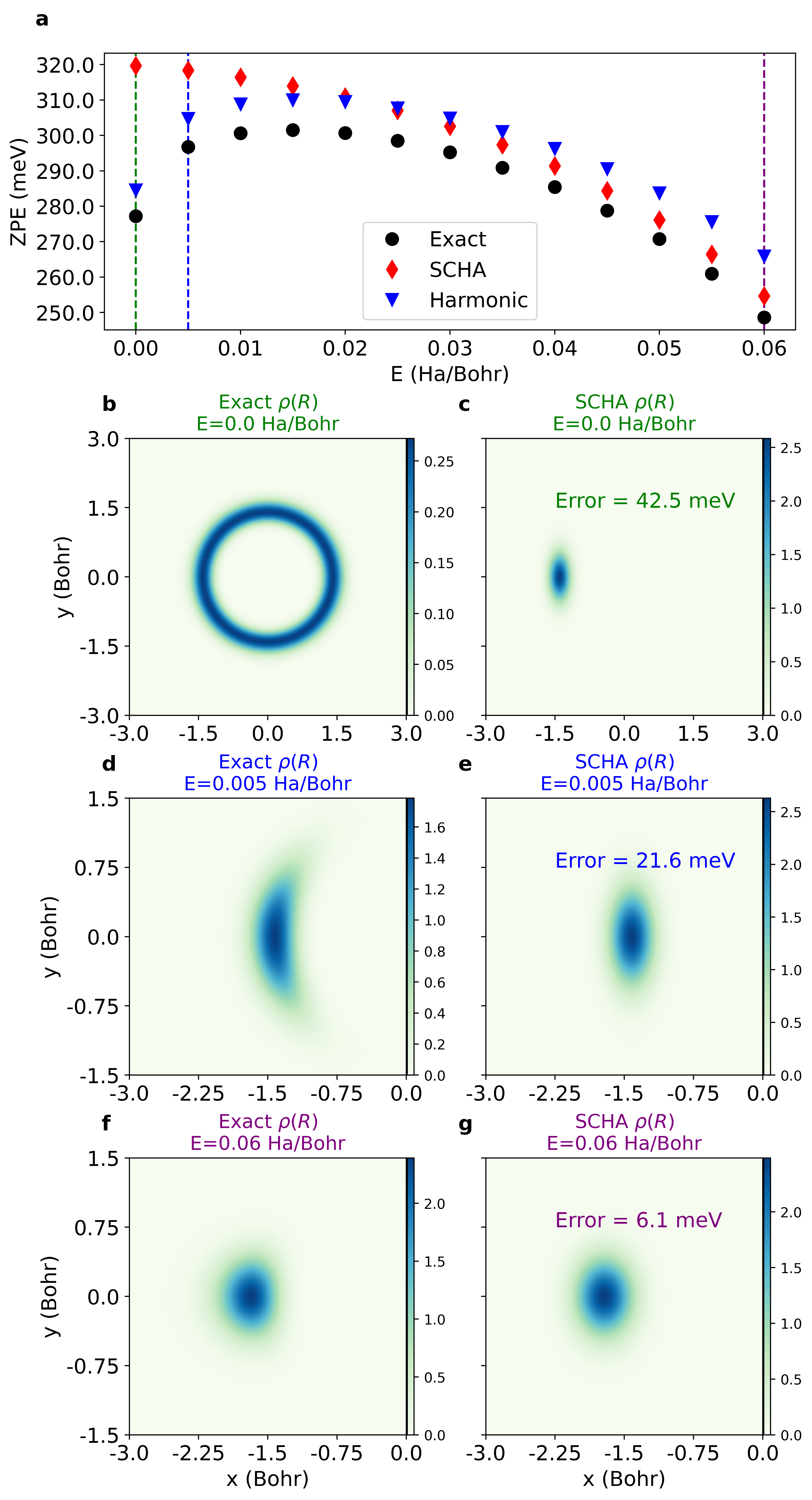}
    \end{minipage}
    \caption{Fig.\ a: exact, harmonic, and SCHA zero-point energy (ZPE), i.e. the difference between the ground state energy and the potential minimum. Figs b-g: exact and SCHA probability distributions (in Bohr$^{-2}$) for three values of the crystal field marked by horizontal lines in the upper panel ($E=0,0.005,0.6$ Ha/Bohr). SCHA can not describe roto-librations as its Gaussian trial wavefunction unphysically hybridizes low-energy rotations with high-energy vibrations.}
    \label{fig:scha vs exact T=0K all E}
\end{figure} 
In Fig.\ \ref{fig:scha vs exact T=0K all E} a, we compare the exact, harmonic, and SCHA zero-point energy (ZPE) as a function of the rotational freedom, $E$. In the harmonic approximation, we expand $V^{\text{(BO)}}(\bm{R})$ around its minimum, and from the second-order terms we extract the normal modes. The details of the exact diagonalization and SCHA simulations are in appendix \ref{APP: Exact diagonalization} and appendix \ref{APP: SCHA simulations}. The ZPE is the difference between the ground state energy and the minimum of the potential, representing the energy excess due to quantum uncertainty. In the harmonic approximation, the ZPE is
\begin{equation}
\label{eq: ZPM harmonic}
    \text{ZPE}_\text{harm} =\frac{\hbar\left(\omega_\text{harm,vib} + \omega_\text{harm,rot}\right)}{2}
\end{equation}
where $\omega_\text{harm,vib}, \omega_\text{harm,rot}$ are the frequencies of the vibrational and rotational modes (respectively polarized along x and y).
In Fig.\ \ref{fig:scha vs exact T=0K all E} b, d, f, we represent the probability distribution of the relative coordinate $\bm{R}$ increasing the value of the crystal fields. In the absence of $E$, the \ch{H2} molecule behaves like a free rotator (Fig.\ \ref{fig:scha vs exact T=0K all E} b), then rotations are progressively suppressed as $E$ increases (Fig.\ \ref{fig:scha vs exact T=0K all E} d) until the molecule is locked with a fixed orientation (Fig.\ \ref{fig:scha vs exact T=0K all E} f).

\begin{figure}[!htb]
    \centering
    \begin{minipage}[c]{1.0\linewidth}
    \includegraphics[width=1.0\textwidth]{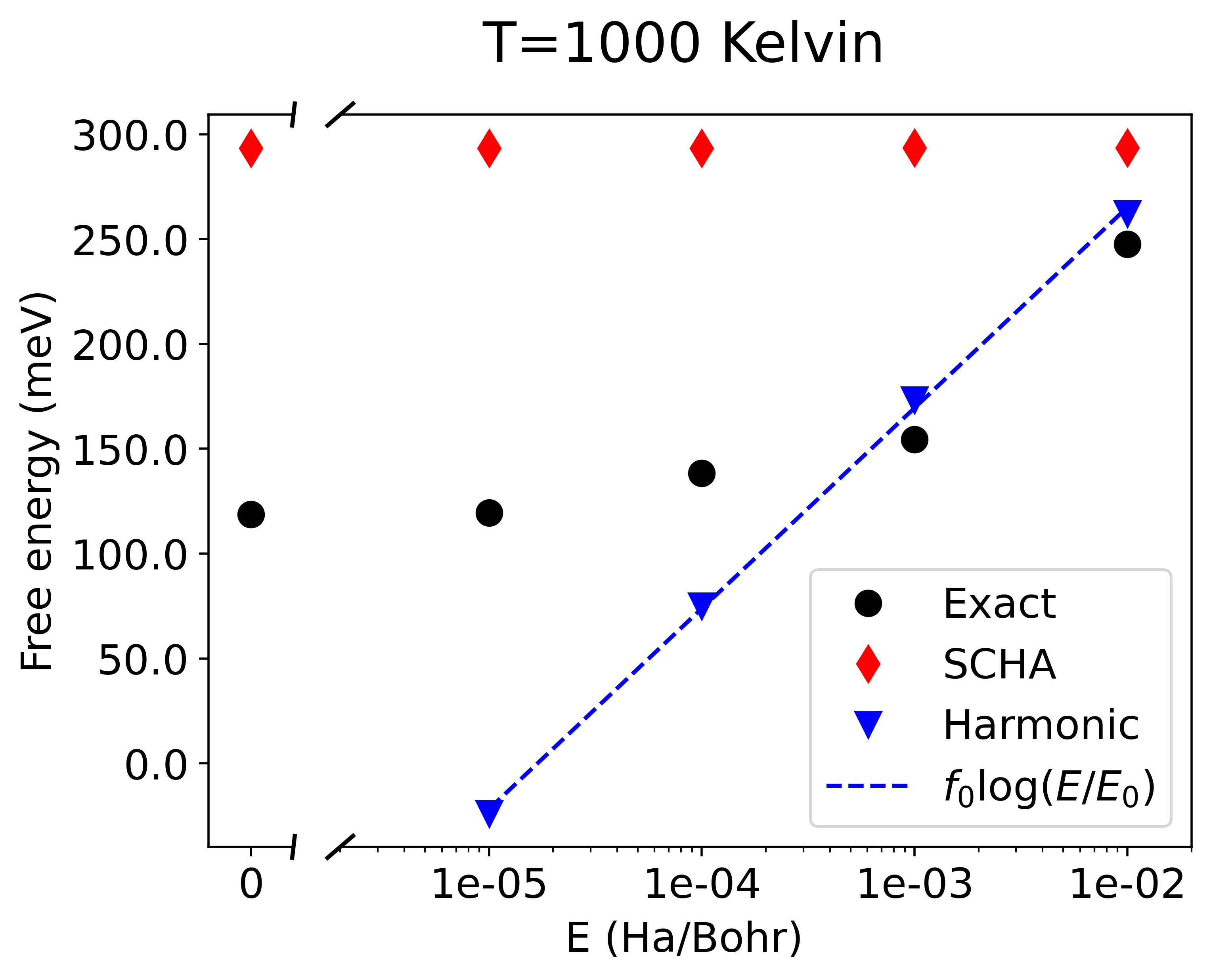}
    \end{minipage}
    \caption{Exact, SCHA and harmonic free energies at $1000$ K for low values of the crystal field $E$. As we reduce $E$, the rotational frequency is vanishing $\omega_\text{harm,rot} \rightarrow 0$, hence the harmonic free energy diverges as $f_0 \log(E/E_0)$ (solid dotted line).}
    \label{fig:fail harmonic}
\end{figure}

In the harmonic approximation, $\omega_\text{harm,rot}$ is zero at $E=0$ Ha/Bohr, indicating the presence of a free rotator mode. So, there is a direction along which the propagation costs zero energy, thereby undermining the assumption of small oscillations on which the HA relies. Consequently, in the case of full rotational invariance, the HA is not justified at finite temperatures. In Fig.\ \ref{fig:fail harmonic}, we compare the exact, SCHA and harmonic free energies at $1000$ K for low values of $E$. In the limit of vanishing crystal field, the harmonic entropy diverges logarithmically as the rotational frequency tends to zero. Consequently, the HA is unreliable at finite temperatures for low values of $E$.

In contrast, the SCHA method effectively handles large nuclear fluctuations by adjusting the width and center of the Gaussian, minimizing the total energy. Thus, we always get well-defined solutions. However, the SCHA method samples the BOES using a Gaussian distribution in Cartesian coordinates, Fig.\ \ref{fig:scha vs exact T=0K all E} c, e, g. The approximation employed by SCHA results in an overestimation of the rotational barrier, as it hybridizes vibrational and rotational modes, causing an unphysical stiffening of the latter. This is evident when comparing SCHA and exact probability distribution,  Fig.\ \ref{fig:scha vs exact T=0K all E} b-e. So a Gaussian trial density matrix lacks the flexibility to describe free or partially rotating molecules, thus sampling high-energy regions of the BOES.
As a result, the SCHA ZPEs are unreliable at low crystal fields. They can be worse than the predictions from the HA, that, however provides a non-variational energy. Therefore, the SCHA method fails to meet the gold-standard free energy method requirements when rotations are present \cite{ceriotti_anharmonic_free_energies,HydrogenIVMorresi}. Nevertheless, as $E$ is increased, rotations are suppressed, and the SCHA method performs better, adeptly capturing the anharmonic effects of the Morse potential and reducing its error in comparison to the exact solution, as seen in Fig.\ \ref{fig:scha vs exact T=0K all E} f-g.

Here, we demonstrated again that the SCHA is a valuable tool when linear vibrations dominate as already noted in Refs \cite{ceriotti_anharmonic_free_energies,HydrogenIVMorresi}. However, as soon as rotational modes become active, the method falters due to the trial density matrix's lack of flexibility.  All the other methods belonging to the SCP family have the same problem. So, SCP methods can not detect the presence of a rotational mode, leading to an overestimation of the total energy. To address these issues, we employ the NLSCHA theory \cite{INPREPARAZIONE}, which adapts a new trial density matrix by automatically activating a roto-librational mode if it lowers the total energy.

\section{The solution of nonlinear Self-Consistent Harmonic Approximation}
\label{SEC: The solution of nonlinear Self-Consistent Harmonic Approximation}
The SCP approach falls short in considering roto-librations, as these modes generate non-Gaussian fluctuations (refer to Fig.\ \ref{fig:scha vs exact T=0K all E} b-d), surpassing the capabilities of these methods. We employ the NLSCHA theory to modify the Gaussian \textit{ansatz} with an invertible nonlinear transformation so that it can accommodate rotations.

\subsection{The nonlinear change of variables}
\label{SUBSEC: The nonlinear change of variables}
We aim to disentangle rotations from stretching modes as the SCHA leads to unphysical hybridization. The solution of NLSCHA is to variationally optimize a harmonic density matrix in a suitable auxiliary space that separates these modes. 
\begin{figure*}[!htb]
    \centering
    \begin{minipage}[c]{1.0\linewidth}
    \includegraphics[width=1.0\textwidth]{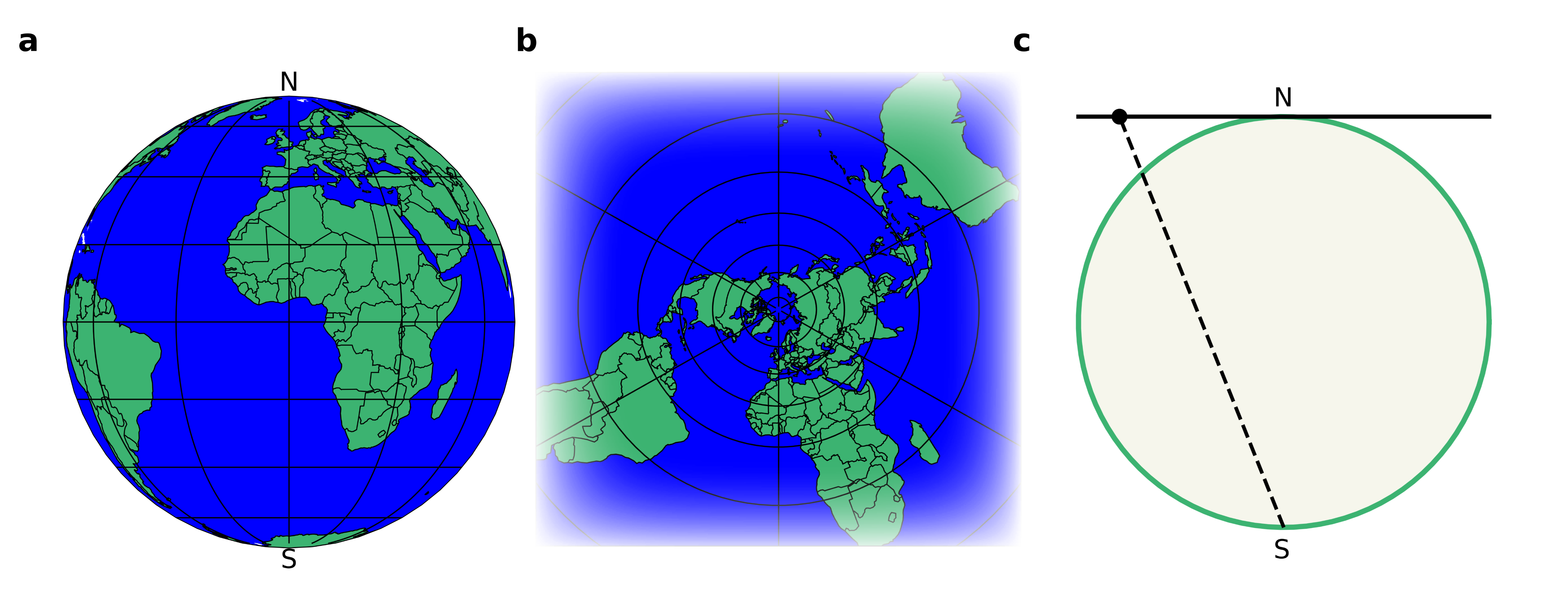}
    \end{minipage}
    \caption{The stereographic transformation represents the Earth's surface (Fig.\ a) on a 2D plane (Fig.\ b). Each point on the Earth, identified by $(\theta,\phi)$, is mapped to a unique point $(x_\text{s}, y_\text{s})$ on a plane tangent to the North Pole (N) (Fig.\ c). To build this transformation, We connect S to a point $(\theta,\phi)$ then the stereographic projection $(x_\text{s}, y_\text{s})$ is given by the intersection of this line with the tangent plane as Fig.\ c shows. This change of variables applied to the atomic coordinates disentangles rotational modes (e.g.\ the motion on the stereographic plane) from the vibrational ones (e.g.\ the motion perpendicular to the surface). The plot was made with the \texttt{cartopy} package \cite{Cartopy}.}
    \label{fig:cartography}
\end{figure*}
The most straightforward choice to describe rotations is spherical coordinates $(r,\theta,\phi)$ \cite{HydrogenIVMorresi}
\begin{subequations}
\label{def: circular coords in 2D}
\begin{align}
    r & = \sqrt{x^2 + y^2 + z^2} \\
    \phi & = \atan\left(\frac{y}{x}\right) \\
    \theta & = \arccos\left(\frac{z}{\sqrt{x^2 + y^2 + z^2}}\right)
\end{align}
\end{subequations}
Note that it is possible to define small oscillations (i.e.\ the harmonic approximation) only in 2D $(r,\phi)$ but not in 3D $(r,\phi,\theta)$ as small variations of the polar angle $\theta$ around $\theta \simeq 0$ lead to large changes of the azimuthal one $\phi$. In addition, Eqs \eqref{def: circular coords in 2D} have a different topology from the Cartesian space, indeed $(r,\phi,\theta)$ are not defined in $\mathbb{R}^3$. To use the NLSCHA framework, we must work with an auxiliary space with the same topology as the Cartesian one so that small oscillations are well-defined and the entropy is an analytical quantity \cite{INPREPARAZIONE}. It is not a problem working with $r$ as it represents the atomic distance and exploring $r\rightarrow 0$ has a low probability. Indeed, $r\rightarrow 0$ has a substantial energetic cost as it would correspond to nuclear fusion. So, we have to solve the problem of the angular variables' bounded range. If we adopt the stereographic projection, the topology of $(\phi,\theta)$ becomes the same as the Cartesian space. This technique represents the Earth on a 2D plane with range $\mathbb{R}^2$. All the Earth's surface points are projected on a tangent plane at the North Pole (N) using the South Pole (S) as the projecting point. Note that we can not represent S in this way and that no invertible transformation can achieve this. 

The NLSCHA auxiliary manifold we define here derives from a parameterization of the stereographic transformation as it allows to explore more phase space, i.e.\ it represents both North and South Hemispheres in cartography. For our purposes, this property implies that we describe large angular fluctuations. To better illustrate our approach, we consider the \ch{H2} molecule in 2D. We define the auxiliary variables $\bm{\unlscha} = (\unlscha_1, \unlscha_2)$ from
\begin{equation}
    \bm{R} = \xibmnlscha(\bm{\unlscha})
\end{equation}
where $\bm{R} = (x,y)$ is the \ch{H2} relative coordinate in the center of mass reference frame and $\xibmnlscha(\bm{\unlscha})$ is the stereographic projection
\begin{subequations}
\label{def: stereographic projection in 2D}
\begin{align}
    x(\bm{\unlscha}) &= \xcnlscha + (\unlscha_1 + \rOnlscha) \cos(\phi(\unlscha_2))\\
    y(\bm{\unlscha}) &= \ycnlscha + (\unlscha_1 + \rOnlscha) \sin(\phi(\unlscha_2)) \\
    \phi(\unlscha_2) &= \phiOnlscha + 2 \atan\left(\frac{\unlscha_2}{2r_0}\right) \label{eq: phi - phi0 nlscha}
\end{align}
\end{subequations}
Note that $\xibmnlscha(\bm{\unlscha})$ is parametrized by $\xcnlscha$, $\ycnlscha$, $\rOnlscha$, and $\phiOnlscha$. They represent the center of curvature, $\RCnlscha$, and the curvature vector, $\RTnlscha$, 
\begin{subequations}
\label{eq: nlscha centroids}
\begin{align}
    \RCnlscha &= \begin{bmatrix}
    \xcnlscha & \ycnlscha
    \end{bmatrix} \label{eq: R_C} \\
    \RTnlscha &=
    \begin{bmatrix}
    \rOnlscha \cos(\phiOnlscha) & \rOnlscha \sin(\phiOnlscha)
    \end{bmatrix} \label{eq: R_T}
\end{align}
\end{subequations}
where $\RTnlscha$ contains the information on the curvature $\kappa$ of the nonlinear transformation
\begin{equation}
\label{def: curvature nlscha}
    \kappa=|\RTnlscha|^{-1} = \rOnlscha^{-1}
\end{equation} 
\begin{figure}[!htb]
    \centering
    \begin{minipage}[c]{1.0\linewidth}
    \includegraphics[width=1.0\textwidth]{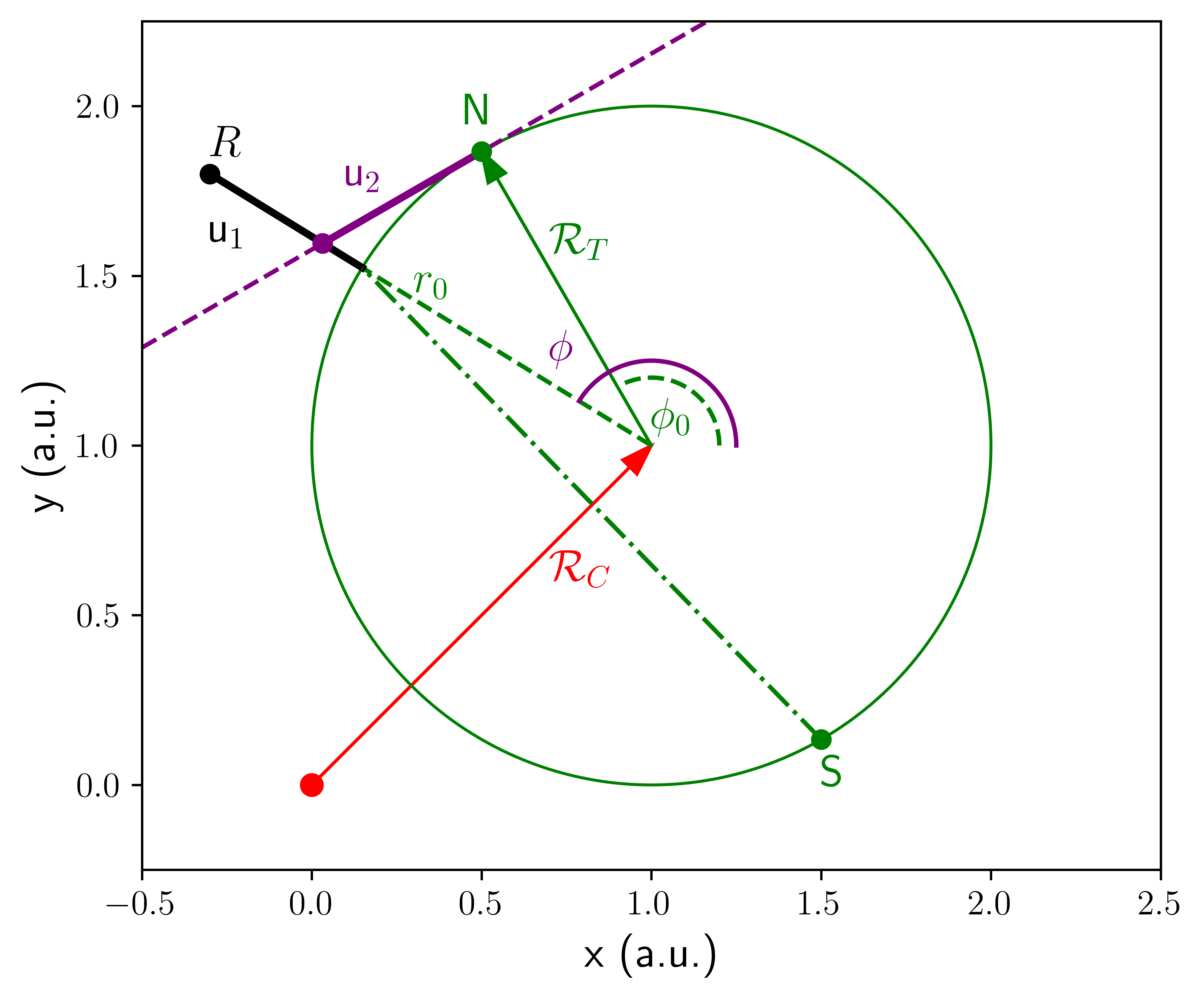}
    \end{minipage}
    \caption{A geometrical representation of the nonlinear transformation, Eq.\ \eqref{def: stereographic projection in 2D}, with curvature $\kappa >0$. $\bm{R} = \xibmnlscha(\bm{\unlscha})$ corresponds to a stereographic projection in 2D. The center and radius of the circle are defined respectively by $\RCnlscha$ (red arrow) and $|\RTnlscha|$ (green arrow). $\RCnlscha + \RTnlscha$ indicates the position of the tangent line, i.e.\ the North Pole. The stereographic projection maps the points $\unlscha_2$ on angles $\phi$, with the South Pole $\RCnlscha -\RTnlscha$ being the projecting point.}
    \label{fig:stereographic 2D}
\end{figure} 
We examine Fig.\ \ref{fig:stereographic 2D} to illustrate the geometrical meaning of Eq.\ \eqref{def: stereographic projection in 2D} with $\kappa>0$. The vector $\RCnlscha$ identifies the circle's center (center of the Earth) with radius $|\RTnlscha|$. $\unlscha_1= |\bm{R} - \RCnlscha| - r_0$, the radial coordinate, and $\unlscha_2$, the stereographic projection of $\phi - \phi_0$, identify the position of a point $\bm{R}=(x,y)$ in the new space. We perform the stereographic projection on the tangent plane at $\RCnlscha + \RTnlscha$ (the North Pole) using $\RCnlscha - \RTnlscha$ (the South Pole) as the projecting point. Note that the North and South Pole are free parameters of the transformation. As $|\RTnlscha|\rightarrow +\infty$, the curvature $\kappa\rightarrow 0$ so we recover a linear transformation (see section \ref{SUBSEC: The density matrix})
\begin{equation}
\label{eq: limit of zero curvature}
    \bm{R} \simeq \RCnlscha - \RTnlscha + \bm{\unlscha} = \bm{\mathcal{R}} + \bm{\unlscha}
\end{equation}
which is the one employed by standard SCHA \cite{SCHA_main} ($\RCnlscha - \RTnlscha=\bm{\mathcal{R}}$ is the average atomic position in SCHA).

\subsection{The trial density matrix}
\label{SUBSEC: The density matrix}
Here, we show how to incorporate the \textit{ad-hoc} change of variables, Eq.\ \eqref{def: stereographic projection in 2D}, with NLSCHA \cite{INPREPARAZIONE}. Within our framework, we variationally minimize the quantum free energy with a trial density matrix corresponding to a Gaussian probability distribution in $\bm{\unlscha}$, not in $\bm{R}$ as in the SCHA. Indeed, depending on $\kappa$, we have a probability distribution that supports roto-librations.
The NLSCHA density operator is defined by the matrix elements (see appendix \ref{APP: The trial density matrix})
\begin{equation}
\label{eq: def rho nlscha R R'}
    \bra{\bm{R}} \rhohatnlscha \ket{\bm{R}'} =  \frac{\gaussnlscha(\bm{\unlscha}, \bm{\unlscha}')}{\sqrt{\detJnlscha(\bm{\unlscha}) \detJnlscha(\bm{\unlscha}')}}
\end{equation}
where $\detJnlscha$ is the Jacobian's determinant 
\begin{equation}
\label{eq: def J dR du}
    \detJnlscha(\bm{\unlscha}) = \det\left(\pdv{\xibmnlscha(\bm{\unlscha)}}{\bm{\unlscha}}\right) > 0
\end{equation}
and $\gaussnlscha(\bm{\unlscha}, \bm{\unlscha}')$ satisfies ($\beta^{-1}=k_\text{B} T$)
\begin{subequations}
\label{eq: def rho nlscha u u'}
\begin{align}
     \gaussnlscha(\bm{\unlscha}, \bm{\unlscha}') & =
    \frac{\bra{\bm{\unlscha}}\exp{-\beta \Hhatnlscha}\ket{\bm{\unlscha}'}}{\Znlscha} \\
     \Znlscha & = \myint d\bm{\unlscha} 
     \bra{\bm{\unlscha}}\exp{-\beta \Hhatnlscha}\ket{\bm{\unlscha}}
\end{align}
\end{subequations}
In NLSCHA the auxiliary harmonic Hamiltonian $\Hhatnlscha$ is defined in $\bm{\unlscha}$-space
\begin{equation}
\label{eq: def H nlscha u u'}
    \bra{\bm{\unlscha}}
    \Hhatnlscha
    \ket{\bm{\unlscha}'} = \delta(\bm{\unlscha}- \bm{\unlscha}')\left(
    -\frac{\hbar^2}{2}
    \pdv{}{\bm{\unlscha}} \cdot \overset{-1}{\masstnsbm} \cdot  
    \pdv{}{\bm{\unlscha}} + \frac{1}{2} \bm{\unlscha} \cdot\FCbmnlscha \cdot \bm{\unlscha}\right)
\end{equation}
The variational parameters of $\rhohatnlscha$ (Eq.\ \eqref{eq: def rho nlscha R R'}) are the force constant $\FCbmnlscha$, the mass tensor $\bm{\mathcal{M}}$ and the free parameters of the nonlinear transformation $\xibmnlscha$
\begin{equation}
    \Rallnlscha = (\RCnlscha, \RTnlscha)
\end{equation}
Note that both $\FCbmnlscha$ and $\bm{\mathcal{M}}$  are symmetric and positive definite. 
\begin{figure}[!htb]
    \centering
    \begin{minipage}[c]{1.0\linewidth}
    \includegraphics[width=1.0\textwidth]{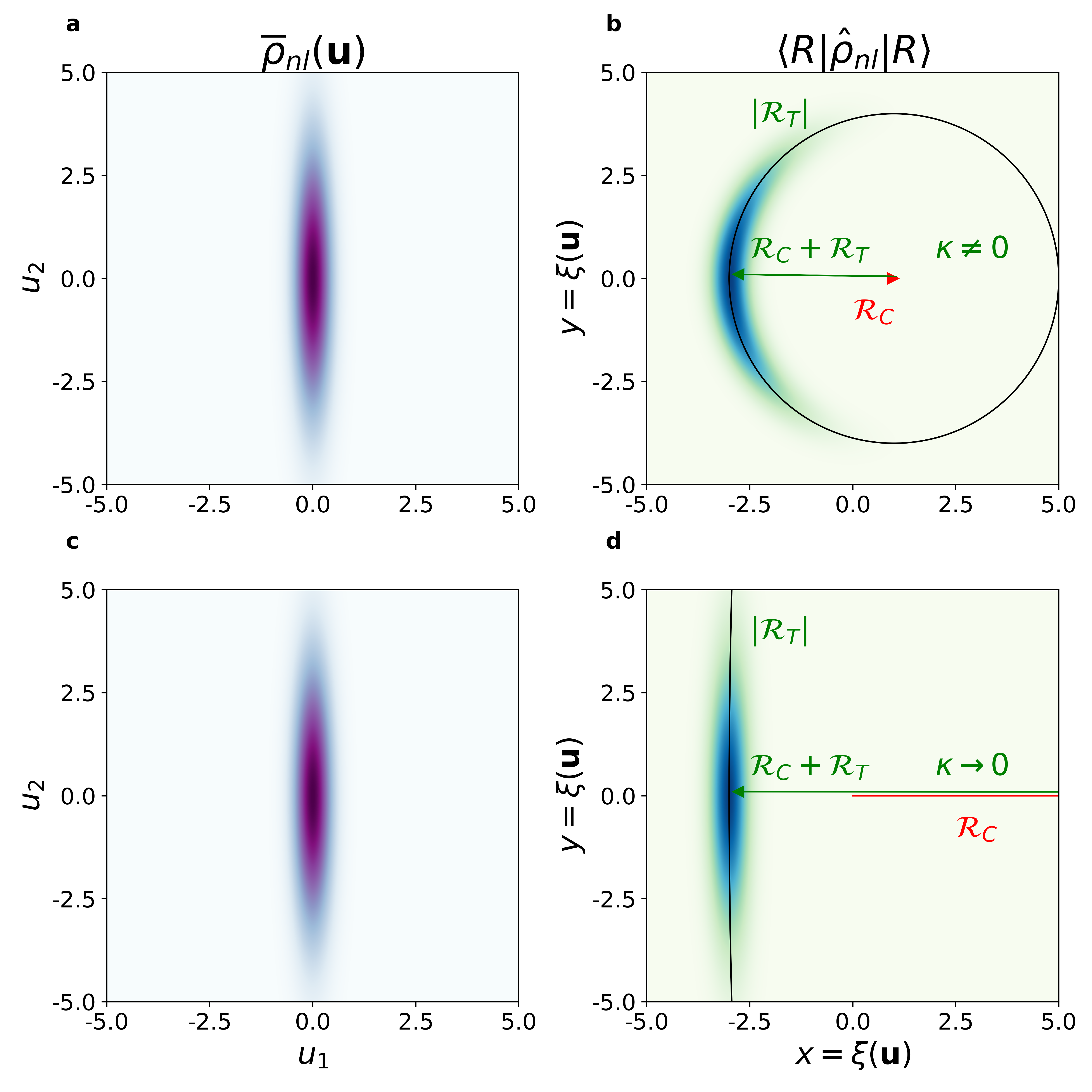}
    \end{minipage}
    \caption{
    Figs a-b show respectively $\gaussnlscha(\bm{\unlscha})$ and the corresponding physical probability distribution $\bra{\bm{R}} \rhohatnlscha \ket{\bm{R}}$ with finite curvature $\kappa=|\RTnlscha|^{-1}>0$. Figs c-d report the same quantities but with zero curvature $\kappa \rightarrow 0$.
    When $\kappa$ is finite, $\bra{\bm{R}} \rhohatnlscha \ket{\bm{R}}$ describes roto-vibrational modes, see Fig.\ b. Otherwise, if $\kappa\rightarrow 0$, the metric of the nonlinear transformation is trivial so $\bra{\bm{R}} \rhohatnlscha \ket{\bm{R}}$ becomes a SCHA-like distribution, see Fig.\ d.}
    \label{fig:rho nlscha 2D}
\end{figure}

We prove that $\bra{\bm{R}} \rhohatnlscha \ket{\bm{R}}$ describes roto-librational modes. In Fig.\ \ref{fig:rho nlscha 2D} we compare $\gaussnlscha(\bm{\unlscha})$, Eqs \eqref{eq: def rho nlscha u u'}, with the corresponding physical probability distribution $\bra{\bm{R}} \rhohatnlscha \ket{\bm{R}}$, Eq.\ \eqref{eq: def  rho nlscha R R'}, showing the effect of the nonlinear transformation when $\kappa >0$ (Fig.\ \ref{fig:rho nlscha 2D} a, b) or $\kappa \rightarrow 0$ (Fig.\ \ref{fig:rho nlscha 2D} c, d). Note that in Fig.\ \ref{fig:rho nlscha 2D}, we keep $\FCbmnlscha$ and $\bm{\mathcal{M}}$ fixed. In the auxiliary $\bm{\unlscha}$-space $\gaussnlscha(\bm{\unlscha})$ is a Gaussian, see Fig.\ \ref{fig:rho nlscha 2D} a-c. When the nonlinear change of variables is applied, the shape of $\bra{\bm{R}} \rhohatnlscha \ket{\bm{R}}$ changes depending on the curvature $\kappa$, see Fig.\ \ref{fig:rho nlscha 2D} b, d. As expected from section \ref{SUBSEC: The nonlinear change of variables}, a finite $\kappa$ bends the probability distribution thanks to the stereographic projection. So $\bra{\bm{R}} \rhohatnlscha \ket{\bm{R}}$ describes a molecule that simultaneously rotates and vibrates, see Fig.\ \ref{fig:rho nlscha 2D} b. On the contrary, when $\kappa \rightarrow 0$, we obtain a Gaussian probability distribution, recovering the standard SCHA, see Fig.\ \ref{fig:rho nlscha 2D} d. 

From Fig.\ \ref{fig:rho nlscha 2D} we deduce the newly introduced variational manifold is a superset of the SCHA, ensuring that NLSCHA systematically outperforms the SCHA for roto-librations. We variationally estimate the exact BO free energy (see appendix \ref{APP: The nonlinear SCHA free energy})
\begin{subequations}
\begin{align}
    F^\text{(BO)} & \leq \Fnl 
    \label{eq: var principle nlscha}\\
     \Fnl &
    = \Tr\left[\rhohatnlscha \hat{H}^\text{(BO)}\right] + k_\text{B} T\Tr\left[\rhohatnlscha\log(\rhohatnlscha)\right] 
    \label{eq: def Fnl}
\end{align}
\end{subequations}
where the entropic term $-k_\text{B}\Tr\left[\rhohatnlscha\log(\rhohatnlscha)\right]$ has a harmonic form and coincides with the temperature derivative of $\Fnl$ only if we optimize all the free parameters \cite{INPREPARAZIONE}. We minimize $\Fnl$ with respect to the free parameters 
\begin{subequations}
\label{eq: nlscha equilibrium conditions}
\begin{align}
    \pdv{\Fnl}{\FCbmnlscha} & = \bm{0} \\
    \pdv{\Fnl}{\bm{\mathcal{M}}} & = \bm{0} \\
    \pdv{\Fnl}{\Rallnlscha} & = 
    \left( \pdv{\Fnl}{\RTnlscha},     
    \pdv{\Fnl}{\RCnlscha} \right) = \bm{0}
\end{align}
\end{subequations}
where the gradient in Eqs \eqref{eq: nlscha equilibrium conditions} depends solely on the BO energies and forces as in the SCHA, see appendix \ref{APPENDIX: Nonlinear SCHA gradient} and Ref.\ \cite{INPREPARAZIONE} for details. 
Once the equilibrium conditions, Eqs \eqref{eq: nlscha equilibrium conditions}, are reached the system's real interacting normal modes are described, in a self-consistent framework, by the NLSCHA auxiliary phonons
\begin{equation}
\label{eq: nlscha phonons}
    \Dbmnlscha = \invsqrtmasstnsTbm \cdot \FCbmnlscha \cdot \invsqrtmasstnsbm
    = \sum_{\mu=1}^{2} 
    \onlscha{\mu}^2 \polbmnlscha{\mu} \polbmnlscha{\mu}
\end{equation}
where $-1$ and $-T$ indicate the inverse and its transpose. The variational optimization of its curvature $\kappa$ allows the description of both linear vibrations (see Fig.\ \ref{fig:rho nlscha 2D} b for $\kappa\rightarrow0$) and roto-librations (see Fig.\ \ref{fig:rho nlscha 2D} a for $\kappa >0$) so the crystal/molecule activates the minimum-free energy degrees of freedom without any external constraint.

\section{Results at zero temperature}
\label{SEC: Results at zero temperature}

\begin{figure}[!htb]
    \centering
    \begin{minipage}[c]{1.0\linewidth}
    \includegraphics[width=1.0\textwidth]{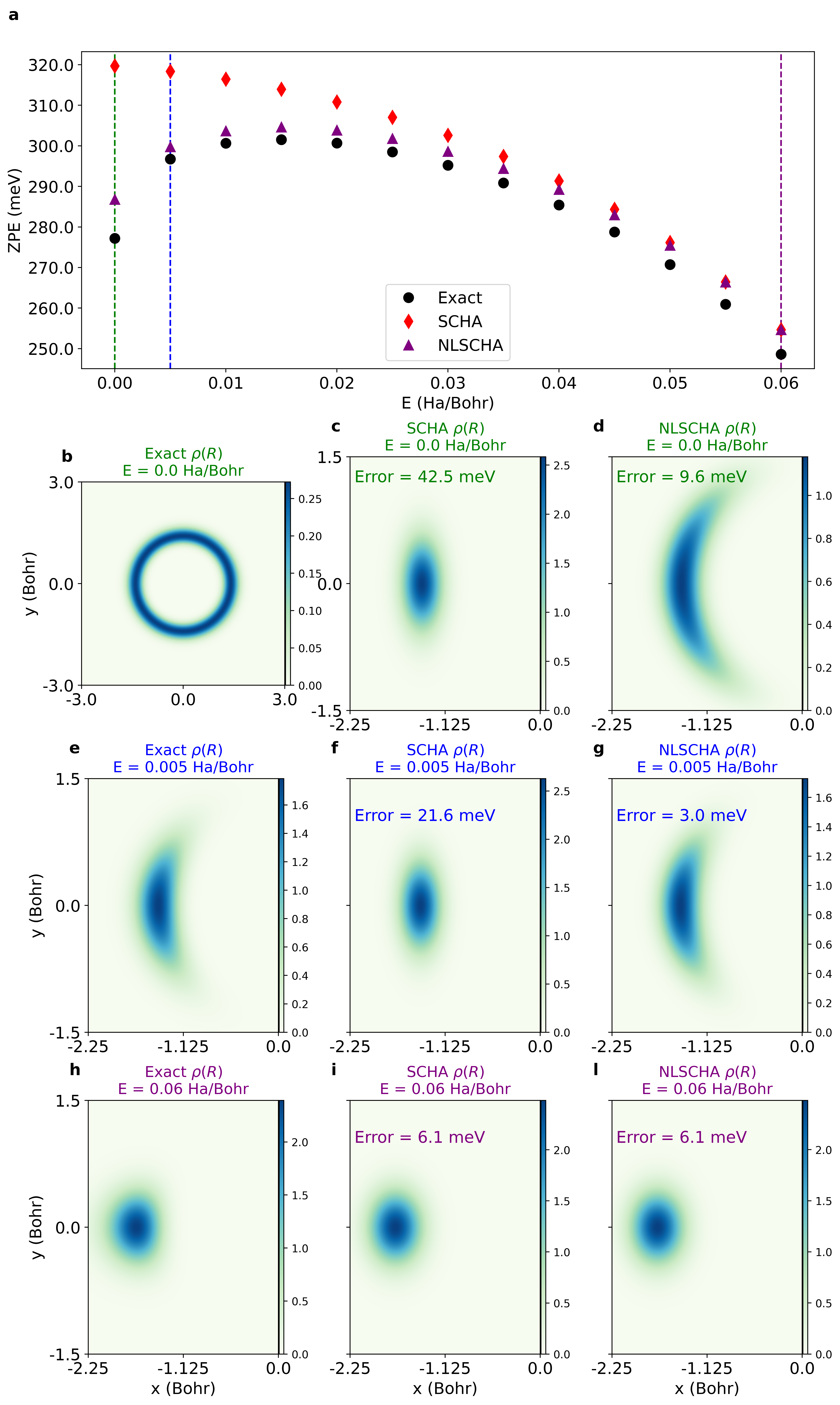}
    \end{minipage}
    \caption{Fig.\ a: exact, SCHA and NLSCHA zero-point energy (ZPE), i.e.\ the difference between the ground state energy and the minimum of the potential. Fig.\ b-l: exact, SCHA, and NLSCHA probability distributions (in Bohr$^{-2}$) for three values of the crystal field marked by horizontal lines $E=0,0.005,0.06$ Ha/Bohr. At low values of $E$, NLSCHA describes roto-librations since the stereographic projection disentangles rotations from vibrations (Figs d-g). At high $E$, NLSCHA reproduces the SCHA results (Figs i-l).}
    \label{fig:scha vs exact vs nlscha T=0K all E}
\end{figure}

The NLSCHA results obtained solving Eqs \eqref{eq: nlscha equilibrium conditions} are reported in Fig.\ \ref{fig:scha vs exact vs nlscha T=0K all E}. In Fig.\ \ref{fig:scha vs exact vs nlscha T=0K all E} a, we report the exact, SCHA and NLSCHA zero-point energy (ZPE) changing the crystal field $E$. Fig.\ \ref{fig:scha vs exact vs nlscha T=0K all E} b-i compare the exact probability distributions with the SCHA and NLSCHA ones for three values of the crystal field $E$.

For low values of $E$, the NLSCHA outperforms the SCHA thanks to the finite curvature $\kappa$ allowing the bending of the vibration in the angular variable. The description of rotations is excellent, in particular, at $E=0.005$ Ha/Bohr (see lower panels of Fig.\ \ref{fig:scha vs exact vs nlscha T=0K all E}) the SCHA error is $21.6$ meV while the NLSCHA error is one order of magnitude smaller, being just $3.0$ meV. 
As already discussed in section \ref{SEC: Failure on rotational modes}, as $E$ is increased, only linear vibrations survive. Note that, reducing the angular fluctuation, NLSCHA perfectly reproduces the SCHA results, Fig.\ \ref{fig:scha vs exact vs nlscha T=0K all E}, see appendix \ref{APP: Nonlinear SCHA simulations}.

\begin{figure}[!htb]
    \centering
    \begin{minipage}[c]{1.0\linewidth}
    \includegraphics[width=1.0\textwidth]{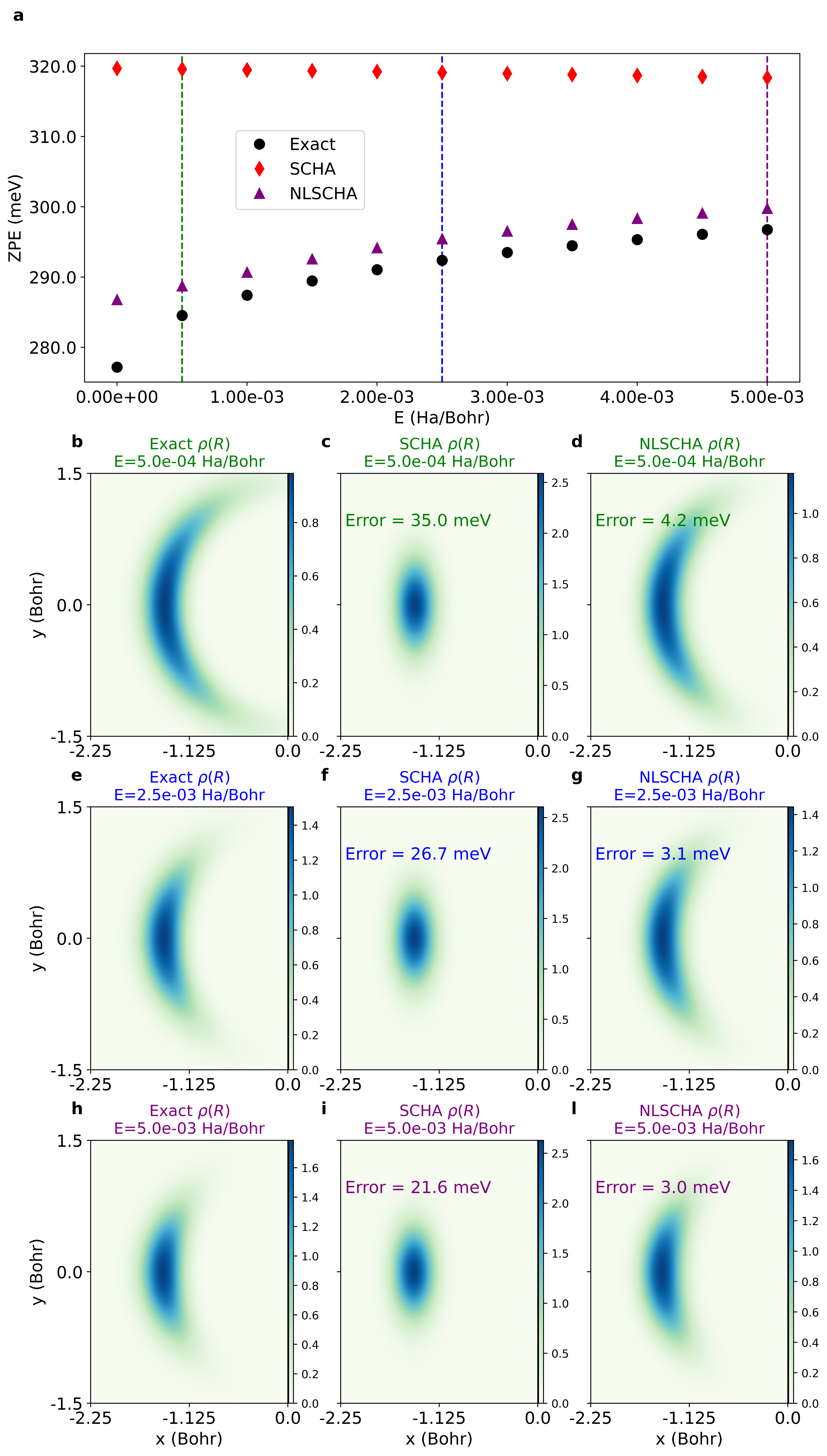}
    \end{minipage}
    \caption{Fig.\ a: exact, SCHA and NLSCHA ZPE for low values of the crystal field (from $E=0$ Ha/Bohr up to $E=5 \cdot 10^{-3}$ Ha/Bohr) which do not suppress roto-librations. Figs b-l: exact, SCHA, and NLSCHA probability distributions (in Bohr$^{-2}$) for three values of the crystal field marked by horizontal lines in the upper panel $E=5 \cdot 10^{-3}, 2.5 \cdot 10^{-3}, 5 \cdot 10^{-3}$ Ha/Bohr. Contrary to SHCA, NLSCHA captures the smallest changes in the rotational degree of freedom thanks to the curvature optimization.}
    \label{fig:scha vs exact vs nlscha T=0K all E zoom}
\end{figure} 

\begin{figure}[!htb]
    \centering
    \begin{minipage}[c]{1.0\linewidth}
    \includegraphics[width=1.0\textwidth]{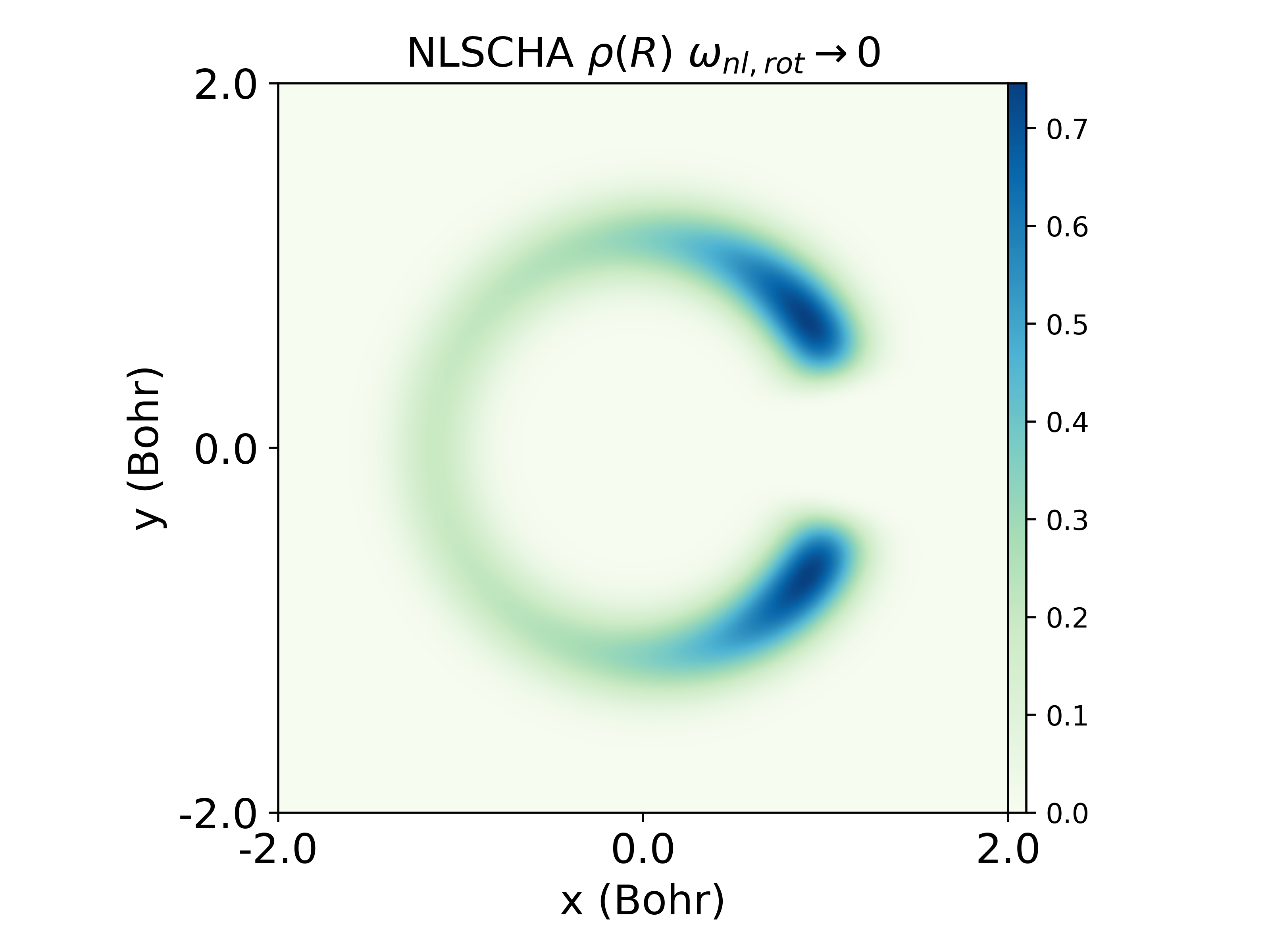}
    \end{minipage}
    \caption{The NLSCHA distribution $\bra{\bm{R}}\rhohatnlscha\ket{\bm{R}}$ (Bohr$^{-2}$) for $\onlscha{\text{rot}}\rightarrow 0$. The limit of vanishing angular frequency, $\onlscha{\text{rot}}\rightarrow 0$, pushes the spectral weight at the boundaries of the stereographic plane so that the NLSCHA Jacobian (Eq.\ \eqref{eq: def J dR du}) deforms the probability distribution. Hence, NLSCHA can not describe free rotations.}
    \label{fig:free rotations NLSCHA}
\end{figure}

In Fig.\ \ref{fig:scha vs exact vs nlscha T=0K all E zoom}, we compare the exact solution with SCHA and NLSCHA for extremely low crystal field values, where the molecule rotates almost freely. Our method detects even the subtlest alterations in the rotational degree of freedom, see Fig.\ \ref{fig:scha vs exact vs nlscha T=0K all E zoom} b-e. Conversely, SCHA is inaccurate as it yields an almost crystal field-independent ZPE.  
While NLSCHA completely captures semi-free rotations (up to $\pi/2$) even for small values of the crystal field, it fails when $E=0$ Ha/Bohr (see Fig.\ \ref{fig:scha vs exact vs nlscha T=0K all E}). Indeed, large angular fluctuations, i.e.\ $\onlscha{\text{rot}}\rightarrow 0$, introduce spurious distortions into the probability distribution preventing a comprehensive characterization of free rotations, see Fig.\ \ref{fig:free rotations NLSCHA}. 

Note that Fig.\ \ref{fig:free rotations NLSCHA} has a connection with cartography. The limit $\onlscha{\text{rot}}\rightarrow 0$ implies diverging angular fluctuations, that push the probability weight at the boundaries of the stereographic plane where the Jacobian (Eq.\ \eqref{eq: def J dR du}) deforms the distribution (Fig.\ \ref{fig:free rotations NLSCHA}). Similarly, the stereographic projection deforms the geographical areas closer to the South Pole (Fig.\ \ref{fig:cartography}). Both in cartography and NLSCHA, this effect is due to the Jacobian (Eq.\ \eqref{eq: def J dR du}) which is dominant for continents below the Equator (Fig.\ \ref{fig:cartography}), and large angular fluctuations (Fig.\ \ref{fig:free rotations NLSCHA}).
Importantly, we emphasize that a multi-peak and sharp distribution as in Fig.\ \ref{fig:free rotations NLSCHA} has a very high kinetic energy, thus making it inaccessible during the minimization of $\Fnl$ (Eq.\ \eqref{eq: def Fnl}). 

\section{Results at finite temperature}
\label{SEC: Results at finite temperature}
\begin{figure}
    \centering
    \begin{minipage}[c]{1.0\linewidth}
    \includegraphics[width=1.0\textwidth]{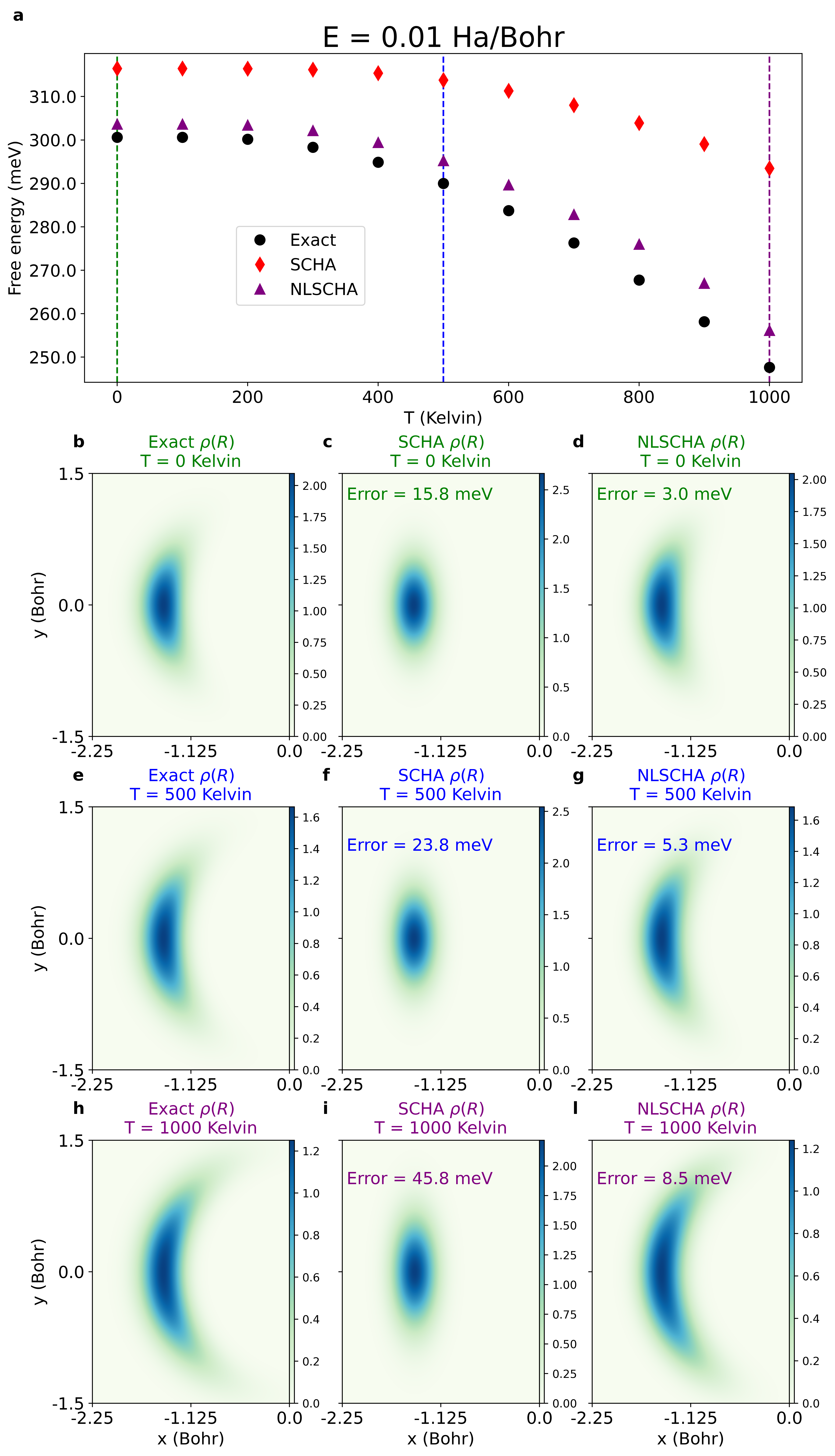}
    \end{minipage}
    \caption{Fig.\ a compares the exact free energy with the SCHA and NLSCHA results ($E=0.01$ Ha/Bohr). Fig.\ b-l: exact, SCHA and NLSCHA probability distribution (in Bohr$^{-2}$) for various temperatures ($0-500-1000$ K). Finite temperature increases the rotational degree of freedom. SCHA completely misses the activation of this degree of freedom whereas in NLSCHA the angular fluctuation is optimized to minimize the free energy variationally.}
    \label{fig:scha vs exact vs nlscha free energy rotational}
\end{figure}
Temperature plays a major role in the thermodynamics of molecules, as it can activate their rotations. Here, we investigate the thermal effect on the \ch{H2} model comparing the SCHA and NLSCHA with exact results.
Fig.\ \ref{fig:scha vs exact vs nlscha free energy rotational} a presents the exact, SCHA, and NLSCHA free energies in the temperature range from $0$ K up to $1000$ K. Furthermore, Figs \ref{fig:scha vs exact vs nlscha free energy rotational} b-l display the exact, SCHA, and NLSCHA probability distributions for $0$, $500$, and $1000$ K.
Figs \ref{fig:scha vs exact vs nlscha free energy rotational} b, c, d show how the amplitude of the rotation increases upon heating. 

Notably, as the temperature rises, the SCHA error also increases as the trial probability distribution does not bend as the NLSCHA one. The NLSCHA approach effectively captures the temperature-induced activation of rotations by appropriately determining the optimal $\kappa$ for each temperature $T$. Notably, at $500$ K, the NLSCHA error is $5.3$ meV while the SCHA completely fails with an error of $23.8$ meV. The NLSCHA error grows with temperature as the entropy is generated by a harmonic Hamiltonian and of course, there is space for future works to relax this hypothesis. 
The free parameters of NLSCHA and SCHA are reported respectively in appendix \ref{APP: Nonlinear SCHA simulations} and appendix \ref{APP: SCHA simulations}.

\section{Phase transition}
\label{SEC: Phase transition}
\begin{figure}
    \centering
    \begin{minipage}[c]{1.0\linewidth}
    \includegraphics[width=1.0\textwidth]{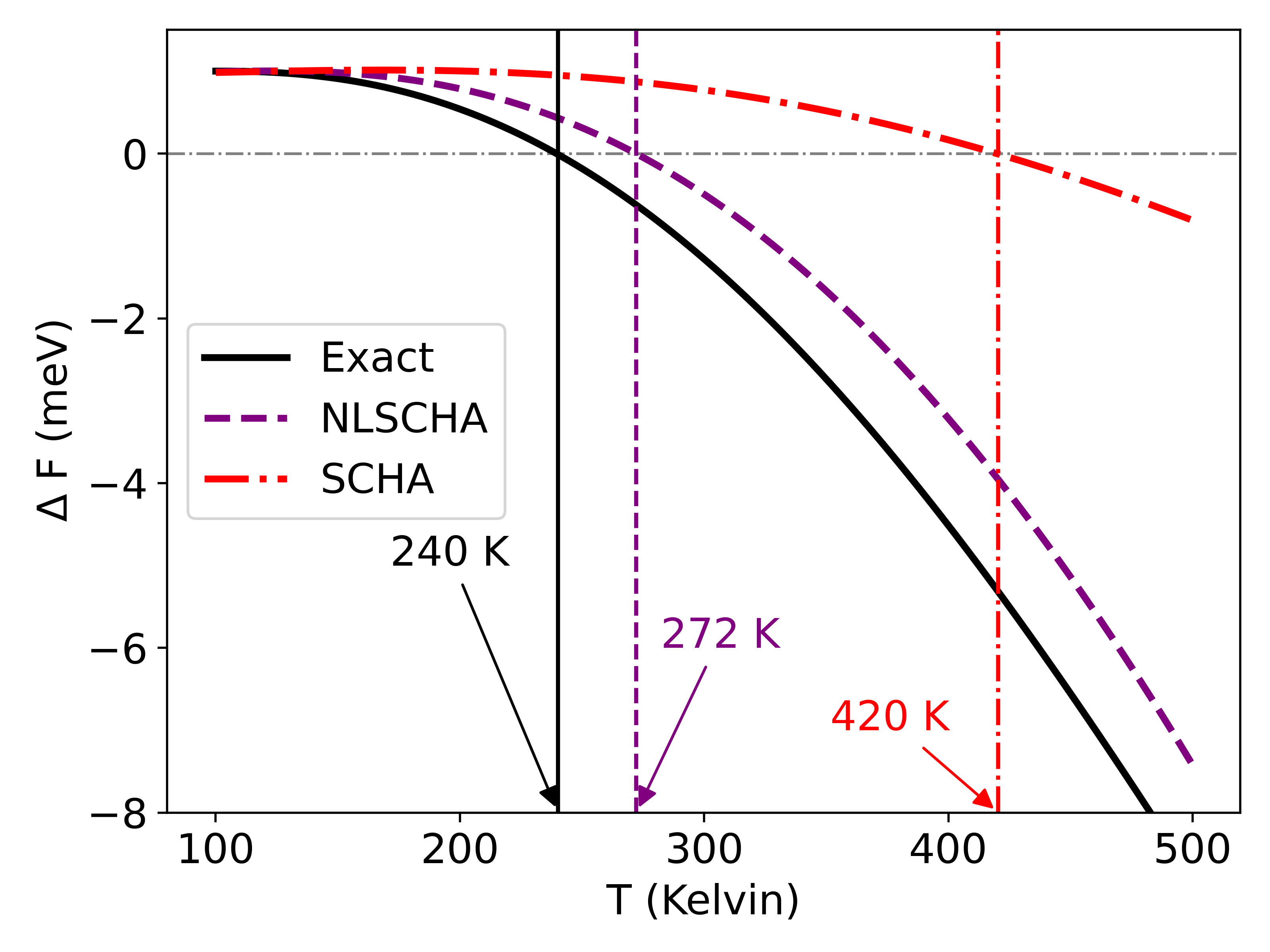}
    \end{minipage}
    \caption{The model for III-IV phase transition of pure hydrogen. Exact, SCHA, and NLSCHA free energy difference $\Delta F = F_\text{rot} - F_\text{vib} + \delta$ as a function of temperature $T$. $F_\text{rot}$ is the free energy for $E=0.01$ Ha/Bohr and
    $F_\text{vib}$ for $E=0.06$ Ha/Bohr.  The critical temperature $T_\text{c}$ (defined by $\Delta F = 0$ meV) depends on the value of the shift $\delta$, which differs for all the methods used. NLSCHA finds the correct critical temperature as it accurately reproduces the temperature-induced activation of rotations (see Fig.\ \ref{fig:scha vs exact vs nlscha free energy rotational}).}
    \label{fig:scha vs exact vs nlscha phase transition}
\end{figure}
Now, we discuss a qualitative model for the III-IV phase transition of high-pressure hydrogen. In phase III, the orientation of the \ch{H2} molecules is fixed \cite{Monacelli_hydrogen_new}, by increasing temperature, \deleted{free} rotations are activated, and phase IV is stabilized at around $300$ K \cite{IV_Gregoryanz1,IV_Gregoryanz2,HydrogenIVMorresi}. 
\added{A particle with reduced mass $\mu_\text{\ch{H2}}$ moving in the potential of Eq.\ \eqref{def: toy model potential} with low values of the crystal field $E$ models a \ch{H2} molecule belonging to phase IV \cite{HydrogenIVMorresi}. Indeed, low $E$ allows for roto-librational modes mimicking the phase IV crystal environment. On the contrary, to model a \ch{H2} molecule in phase III we need high $E$ so that the molecules can only vibrate. Indeed, where phase III is stable, at low temperatures and above $150$ GPa, the intermolecular interactions suppress rotational degrees of freedom \cite{Monacelli_hydrogen_new}.} 

\deleted{We consider the exact, SCHA and NLSCHA free energy difference for two values of the crystal field $E=0.01$, $0.06$  Ha/Bohr. $E=0.01$ Ha/Bohr models phase IV as rotations are thermally activated (see Fig.\ \ref{fig:scha vs exact vs nlscha free energy rotational}), while $E=0.06$ Ha/Bohr represents phase III as these modes are always locked (see appendix \ref{APP: Phase transition model}).} 
\replaced{We denote the phase IV model free energy as $F_\text{rot}$, obtained by setting $E = 0.01$ Ha/Bohr, while $F_\text{vib}$ represents the phase III model free energy, corresponding to $E = 0.06$ Ha/Bohr (see appendix \ref{APP: Phase transition model}). Thus, the quantity $\Delta F = F_\text{rot} - F_\text{vib} + \delta$ ($\delta$ is a different shift for all the methods) represents the phase transition between a structure that does not host roto-librations (phase III) and another one that does (phase IV), see Fig.\ \ref{fig:scha vs exact vs nlscha phase transition}.}{ Fig.\ \ref{fig:scha vs exact vs nlscha phase transition} shows $\Delta F = F_\text{rot} - F_\text{vib} + \delta$ where $\delta$ is a different shift for all the methods.} We define the critical temperature $T_\text{c}$ for each method as the temperature where $\Delta F = 0$ meV. Remarkably, the NLSCHA $T_\text{c}$ is very close to the exact one with an error of $14\%$; on the contrary, in the SCHA, there is an overestimation of \replaced{$75\%$}{$75\%$ K} due to the unphysical hybridization of the rotations with the linear vibrations. 
Hence, we expect SCHA to be completely unreliable for investigating high-pressure hydrogen at finite temperatures and should be replaced by the more powerful NLSCHA. Remarkably, as in the SCHA, the entropy of NLSCHA does not need any additional complex calculation, contrary to MD/PIMD. Indeed, it is analytical and depends solely on the NLSCHA phonons (Eq.\ \eqref{eq: nlscha phonons}) \cite{INPREPARAZIONE}.

\section{3D case}
\label{SECTION: 3D case}
\begin{figure}
    \centering
    \begin{minipage}[c]{0.6\linewidth}
    \includegraphics[width=1.0\textwidth]{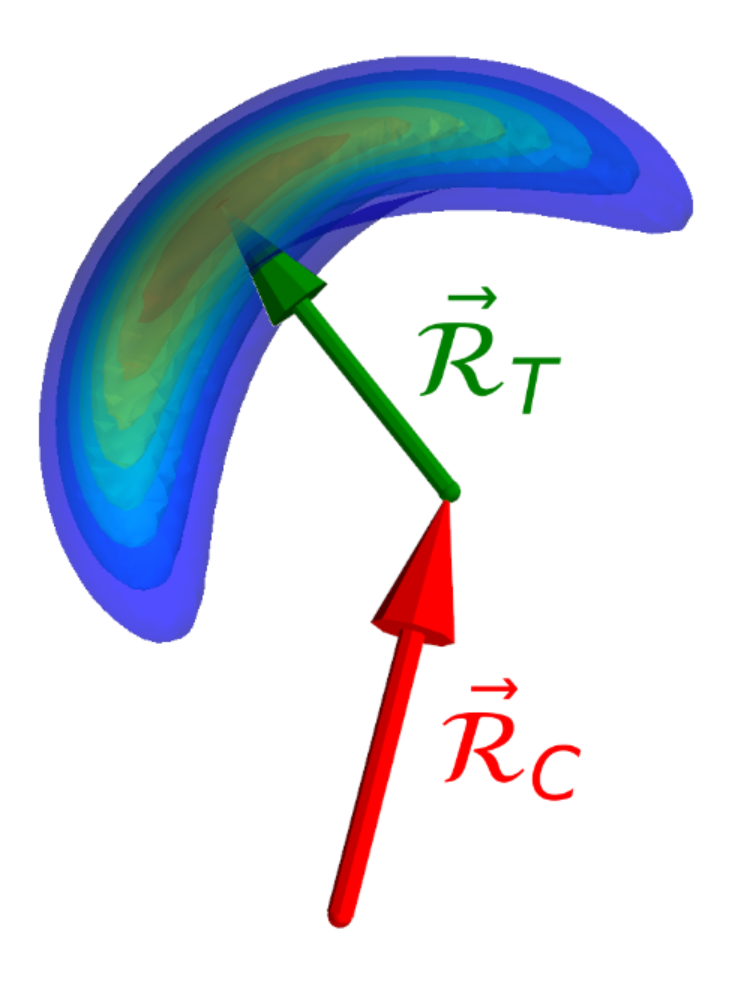}
    \end{minipage}
    \caption{3D isosurfaces of the NLSCHA probability distribution for a diatomic molecule's center of mass $\bm{R}=\bm{R}_1 - \bm{R}_2$. The plot was made with \texttt{mayavi} package \cite{mayavi}.}
    \label{fig:3D case}
\end{figure}

In this section, we extend the nonlinear change of variables discussed in section \ref{SUBSEC: The nonlinear change of variables} to the case of $N$ atoms in three dimensions. Here, we employ the stereographic projection on the sphere, so we transform the Cartesian coordinates $\bm{R}_i$ of each atom $i$ in the following way
\begin{subequations}
\begin{align}
    & \bm{R}_i  = \bm{\mathcal{R}}_{\text{C},i} +   (\unlscha_{i,1} + r_{0,i}) \begin{bmatrix}
    \cos(\phi(\bm{\unlscha}_i)) \sin(\theta(\bm{\unlscha}_i)) \\
    \sin(\phi(\bm{\unlscha}_i)) \sin(\theta(\bm{\unlscha}_i)) \\
    \cos(\theta(\bm{\unlscha}_i))
    \end{bmatrix} \\
    & \phi(\bm{\unlscha}_i)  = \phi_{0,i} + \atan\left(\frac{\unlscha_{i,2}}{\unlscha_{i,3}}\right)\label{eq: 3D nonlinear phi}\\
    & \theta(\bm{\unlscha}_i)  = \theta_{0,i} + 2\atan\left(\frac{\sqrt{\unlscha_{i,2}^2+ \unlscha_{i,3}^2}}{2 r_{0,i}}\right)\label{eq: 3D nonlinear theta}
\end{align}
\end{subequations}
As in section \ref{SUBSEC: The nonlinear change of variables}, the free parameters are the center of the curvature, $\bm{\mathcal{R}}_{\text{C},i}$, and the curvature vector, $\bm{\mathcal{R}}_{\text{T},i}$, defined as
\begin{equation}
    \bm{\mathcal{R}}_{\text{T},i} = r_{0,i} \hspace{-0.1cm}
    \begin{bmatrix}
        \cos(\phi_{0,i})\sin(\theta_{0,i}) & 
        \sin(\phi_{0,i})\sin(\theta_{0,i}) & \cos(\theta_{0,i})
    \end{bmatrix}
\end{equation}
Note that the nonlinear transformation does not mix the coordinates of different atoms; consequently, the number of extra free parameters scales linearly with $N$. $\bm{\mathcal{R}}_{\text{C},i}$ is a linear shift of the positions. $\bm{\mathcal{R}}_{\text{T},i}$ defines the position of the stereographic plane and its length gives the inverse curvature on each atom $r_{0,i}=|\bm{\mathcal{R}}_{\text{T},i}|=\kappa_i^{-1}$. So $\kappa_i > 0$ means that atom $i$ is part of a group that rotates (e.g. an organic molecule inside the cage of a molecular perovskite) otherwise, if $\kappa_i \rightarrow 0$, the atom only vibrates (e.g. the atoms of a cage in a molecular perovskite). Note that we can not use $(\phi,\theta)$ in NLSCHA as small oscillations are not well-defined. Indeed, $\theta\simeq 0$ imply large $\phi$ fluctuations. On the contrary, we always define the harmonic approximation in the stereographic coordinates (see Fig.\ \ref{fig:cartography}).
In Fig.\ \ref{fig:3D case}, we present the NLSCHA probability distribution for a 3D rotating diatomic molecule in the center-of-mass reference frame $\bm{R}=\bm{R}_1 - \bm{R}_2$. Fig.\ \ref{fig:3D case} highlights the applicability of our approach to the study of extended crystals and molecules.

\added{To illustrate better how NLSCHA works in 3D, we briefly discuss how to initialize the nonlinear transformation's parameters in real molecular crystals. For example, we consider pure hydrogen's molecular phases I and II. In these structures, the \ch{H2} centers form a hcp crystal and the molecules show roto-librational modes. Similarly to our 2D model, the vectors $\RCnlscha$ of two atoms forming a \ch{H2} unit will coincide with the molecule's center of mass, while $\RTnlscha$ will connect each $\RCnlscha$ to the corresponding atom. Using this approach to initialize $\RCnlscha$ and $\RTnlscha$, and then optimizing all the free parameters, we would compute the free energy of possible candidate structures, including roto-librational modes, providing reliable phase diagram predictions.}

\added{NLSCHA could also be useful to study the impact of methyl group (\ch{CH3}) roto-librational modes on the stability of biological compounds. \ch{CH3} has a tetrahedral rigid structure and, if its interaction with the environment is weak, the face containing the three hydrogen atoms can rotate around the tetrahedron's height passing through the \ch{C} atom. For example, Ref.\ \cite{ceriotti_anharmonic_free_energies} shows that many approximate methods do not properly describe the \ch{CH3} dynamics in paracetamol, thus failing in finding the correct stable structure at ambient conditions. NLSCHA could capture the rotational degrees of freedom using the stereographic transformation. By symmetry, the center of the curvature $\RCnlscha$ of each \ch{H} atom will be parallel to the tetrahedron's height passing through the \ch{C} atom and will be orthogonal to the face containing the \ch{H} atoms. Then, thanks to the tetrahedral symmetry, the curvature vector $\RTnlscha$ will connect $\RCnlscha$ to each \ch{H} atom.}

\added{Another possible application of NLSCHA is to study the RUMs of network material formed by connected tetrahedra (\ch{SiO4}) or octahedra (\ch{TiO6}). In these structures, for the atoms involved in the rotational modes, $\RCnlscha$ will coincide with the center of the molecular unit while $\RTnlscha$ will point from $\RCnlscha$ to the atom itself satisfying the symmetry of the polyhedra.}

\section{Conclusions}
We have demonstrated that NLSCHA is a very promising method for a systematic and unbiased investigation of molecular crystals' phase diagrams as it takes into account linear vibrations with the same accuracy as SCHA. Still, also it perfectly describes roto-librations, as no other currently available approximation. The improved flexibility is due to the curvature $\kappa$, which minimizes the free energy, allowing the system to spontaneously activate its degrees of freedom.

It is important to emphasize that the additional computational overhead of NLSCHA is negligible. The SCHA and NLSCHA minimize the free energy with respect to an auxiliary force constant matrix, but in NLSCHA we also have the mass tensor $\masstnsbm$. The dimension of these matrices increases quadratically with the number of atoms, thereby being the most computationally intensive step in the minimization. However, while the SCHA minimizes the centroid position $\bm{\mathcal{R}}$, representing the average atomic position, the NLSCHA optimizes two centroids $\RTnlscha$ and $\RCnlscha$ for each atom. The dual centroid structure arises due to the nonlinear transformation and encapsulates information about the particle's average position and the curvature $\kappa_i$. 
Of particular significance is that the nonlinear transformation acts solely on single-particle coordinates; as such, the coordinates of different atoms remain separate (section \ref{SECTION: 3D case}). Hence, the dimensions of $\RTnlscha$ and $\RCnlscha$ scale linearly with the number of atoms. Consequently, the NLSCHA computational cost in the thermodynamic limit scales quadratically with the number of atoms as in the SCHA. This is primarily due to the dominating influence of the optimization process concerning the auxiliary force constant matrix and the mass tensor. 

In conclusion, the NLSCHA equations can be solved stochastically as discussed in Refs \cite{INPREPARAZIONE,SCHA_main}. This makes our method the most promising competitor of PIMD as it incorporates both vibrations and roto-librational degrees of freedom in a quantum framework, finally solving SCHA weakness pointed out in Refs \cite{ceriotti_anharmonic_free_energies,HydrogenIVMorresi}.

\section*{Acknowledgements}
A.S.\ and F.M.\ acknowledge support from European Union
under project ERC-SYN MORE-TEM (grant agreement
No 951215). L.M.\ acknowledges funding from the European Research Council, Marie Curie, project THERMOH.

\appendix

\section{Exact diagonalization}
\label{APP: Exact diagonalization}
\begin{figure}[!htb]
	\centering
	\begin{minipage}[c]{0.7\linewidth}
		\includegraphics[width=1.0\textwidth]{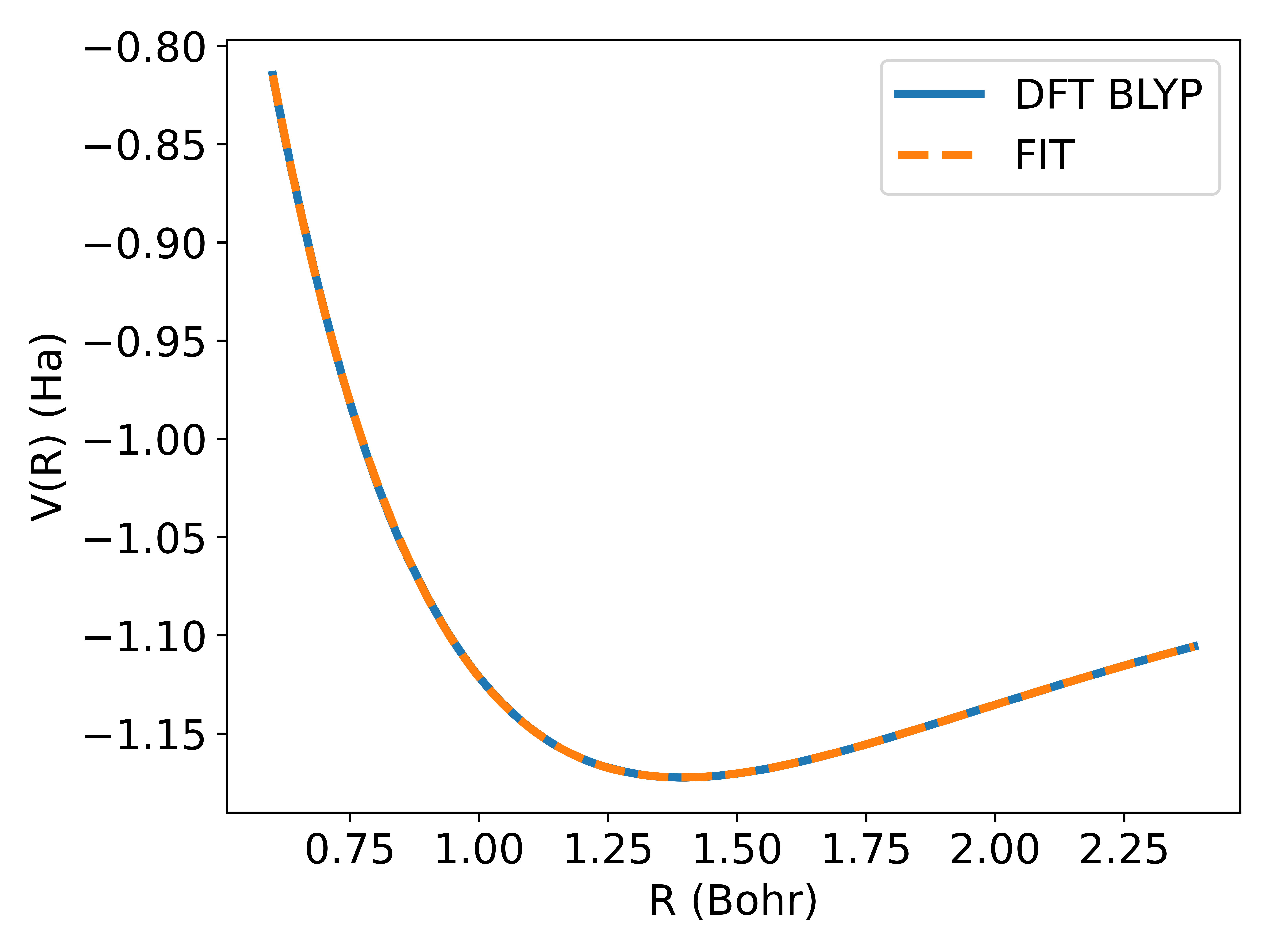}
	\end{minipage}
	\caption{Fit of DFT-BLYP energy profile for \ch{H2} obtained with Quantum Espresso.}
	\label{fig: fit DFT}
\end{figure}

The Morse potential for \ch{H2} is fitted using Quantum Espresso \cite{Giannozzi2009,Giannozzi2017} combined with the BLYP exchange functional \cite{BLYP}, see Fig.\ \ref{fig: fit DFT}.  The cutoffs on plane waves and charge density are $80$ Ry and $320$ Ry, the k-point grid is $(10,10,10)$. The simulation box has size $20$ Bohr.

The exact diagonalization of the Schrodinger equation is performed on a uniform square grid in $\bm{R}$ space of size $N=1200$ between $\pm 3$ Bohr. To get eigenvectors and eigenfunctions of $\hat{H}^\text{(BO)}$, we use the  Implicitly Restarted Lanczos Method \cite{ARPACK} as implemented in the \texttt{scipy} function \texttt{scipy.sparse.linalg.eigsh} \cite{scipy}. The Lanczos algorithm applies many times the target Hamiltonian $\hat{H}^\text{(BO)}$ to a starting normalized wavefunction $\ket{\psi}_\text{init}$ to get the ground state and the first excited states. To avoid the storage of the Hamiltonian as a $N^2 \times N^2$ sparse matrix we used \texttt{scipy.sparse.linalg.LinearOperator} which allows to compute on-the-fly $\hat{H}^\text{(BO)}\ket{\psi}$ where $\bra{\bm{R}}\ket{\psi}$ is defined on the 2D $N\times N$ grid. 

\section{SCHA simulations}
\label{APP: SCHA simulations}
The SCHA simulations were performed on a uniform square grid in $\bm{R}$ space of size $1200$ between $\pm 3$ Bohr. The conjugate gradient (CG) minimization of the SCHA free energy was performed with the \texttt{scipy} \cite{scipy} function \texttt{scipy.optimize.minimize} setting \texttt{gtol} to $10^{-9}$ and \texttt{maxiter} $400$. We report the SCHA free parameters' values both at zero (appendix \ref{APP: SCHA zero temperature}) and finite temperature (appendix \ref{APP: SCHA finite temperature}).

\subsection{Zero temperature}
\label{APP: SCHA zero temperature}
\begin{figure}[!htb]
    \centering
    \begin{minipage}[c]{1.0\linewidth}
    \includegraphics[width=1.0\textwidth]{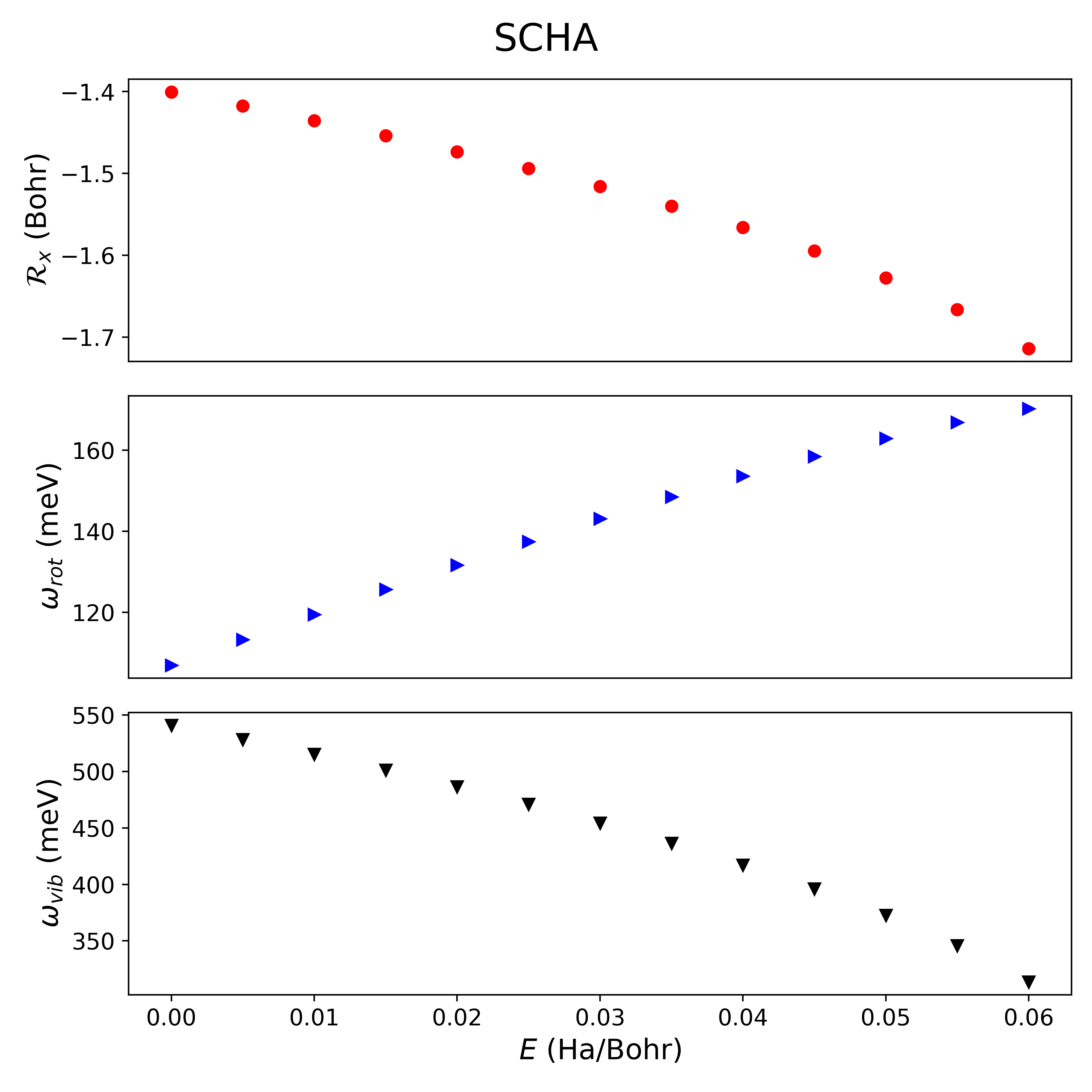}
    \end{minipage}
    \caption{The free parameters of SCHA at zero temperature from $E=0$ Ha/Bohr to $E=0.06$ Ha/Bohr.}
    \label{fig:parameters scha zero T allE}
\end{figure}

\begin{figure}[!htb]
    \centering
    \begin{minipage}[c]{1.0\linewidth}
    \includegraphics[width=1.0\textwidth]{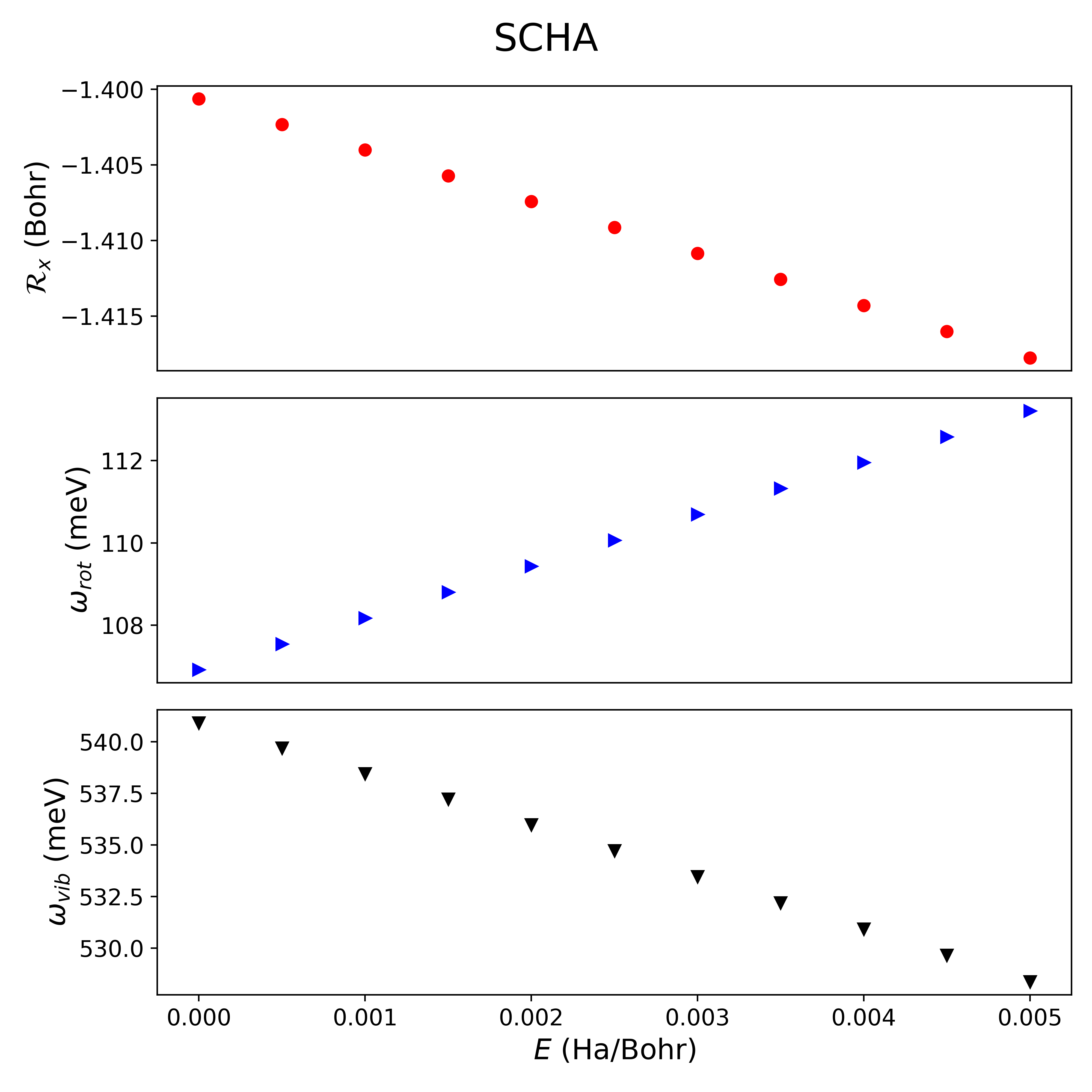}
    \end{minipage}
    \caption{The free parameters of SCHA at zero temperature from $E=0$ Ha/Bohr to $E=0.005$ Ha/Bohr.}
    \label{fig:parameters scha zero T allE zoom}
\end{figure}

In Fig.\ \ref{fig:parameters scha zero T allE} we report the SCHA free parameters at equilibrium from $E=0$ Ha/Bohr to $E=0.06$ Ha/Bohr. $\mathcal{R}_x$ represents the $x$ component of the SCHA centroid, $\bm{\mathcal{R}}$, while $\omega_\text{rot}$ and $\omega_\text{vib}$ are the rotational and vibrational SCHA frequencies. We remark that $\mathcal{R}_y$ is zero by symmetry.
In Fig.\ \ref{fig:parameters scha zero T allE zoom} we report the SCHA free parameters at equilibrium at low values of the crystal field from $E=0$ Ha/Bohr to $E=0.005$ Ha/Bohr.

\subsection{Finite temperature}
\begin{figure}[!htb]
    \centering
    \begin{minipage}[c]{1.0\linewidth}
    \includegraphics[width=1.0\textwidth]{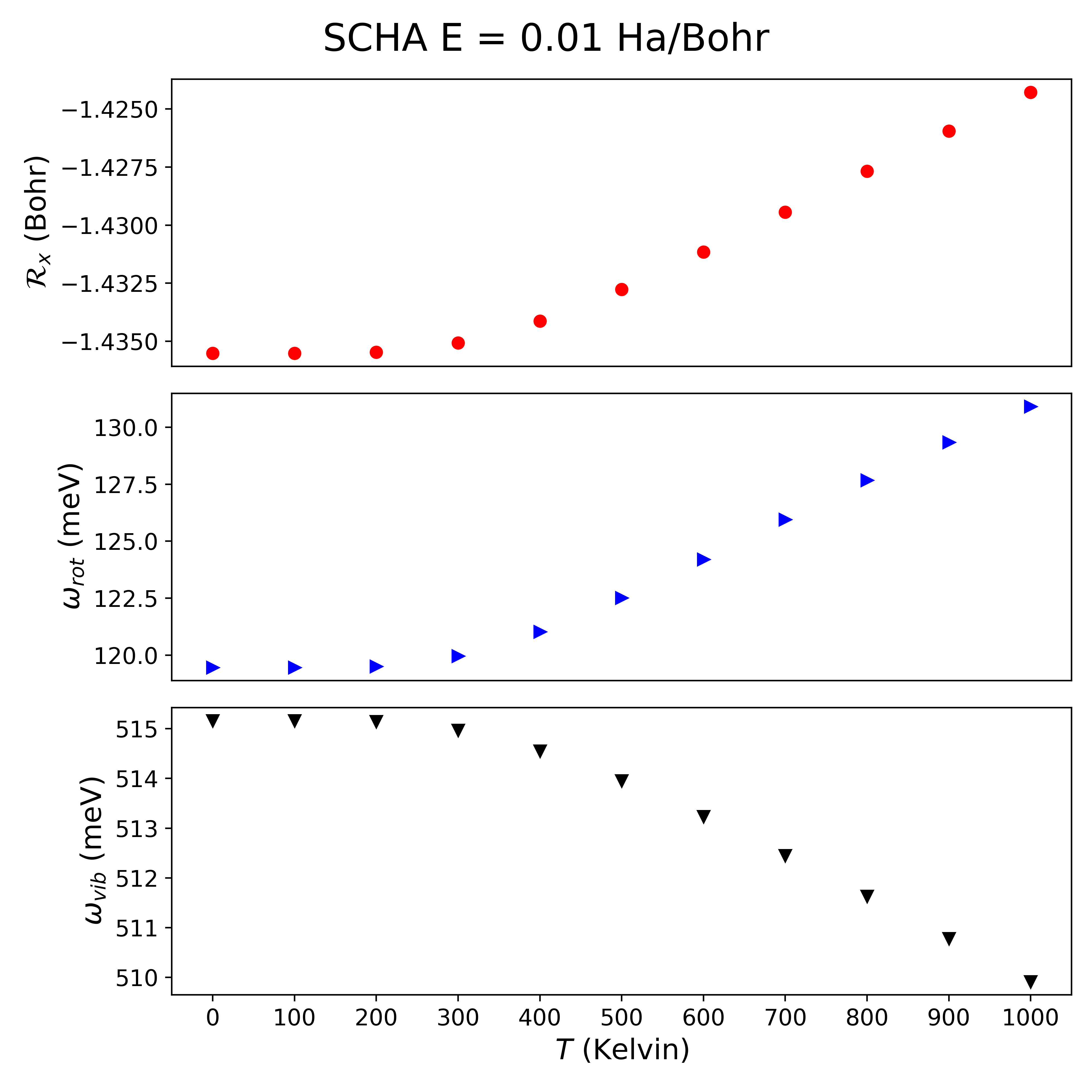}
    \end{minipage}
    \caption{The free parameters of SCHA at finite temperature for $E=0.01$ Ha/Bohr.}
    \label{fig:parameters scha finite T rot}
\end{figure}

\begin{figure}[!htb]
    \centering
    \begin{minipage}[c]{1.0\linewidth}
    \includegraphics[width=1.0\textwidth]{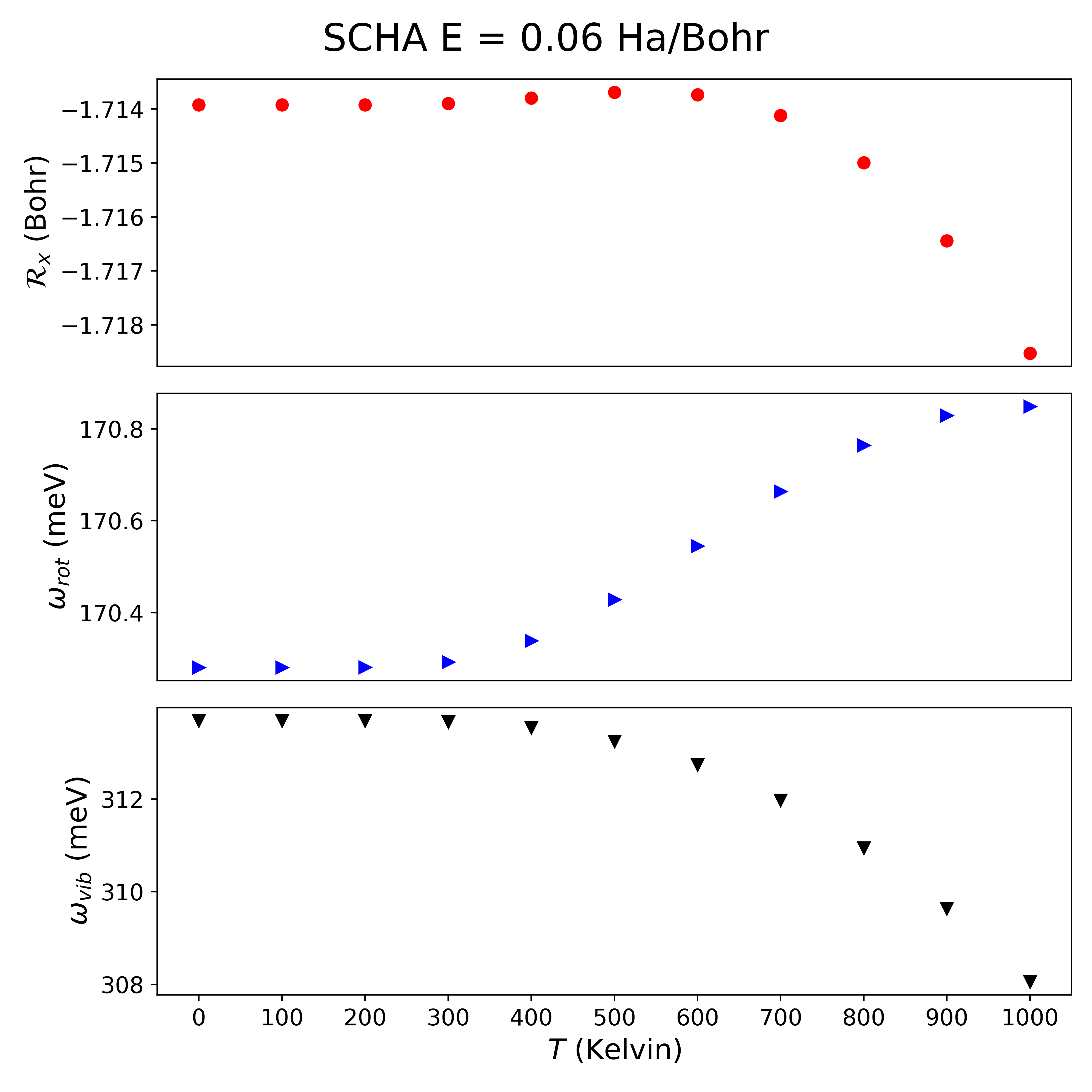}
    \end{minipage}
    \caption{The free parameters of SCHA at finite temperature for $E=0.06$ Ha/Bohr.}
    \label{fig:parameters scha finite T vib}
\end{figure}

The free parameters $\mathcal{R}_x$, $\omega_\text{rot}$ and $\omega_\text{vib}$ at finite temperature for $E=0.01-0.06$ Ha/Bohr are reported in Figs \ref{fig:parameters scha finite T rot} \ref{fig:parameters scha finite T vib}.
\label{APP: SCHA finite temperature}

\newpage

\section{The trial density matrix}
\label{APP: The trial density matrix}
According to Ref.\ \cite{INPREPARAZIONE}, the nonlinear SCHA trial density matrix is given by Eq.\ \eqref{eq: def rho nlscha R R'} where $\gaussnlscha(\bm{\unlscha},\bm{\unlscha}')$ is generated by an harmonic Hamiltonian Eq.\ \eqref{eq: def H nlscha u u'}
\begin{equation}
\label{eq app: rho u u' gauss}
\begin{aligned}
    & \gaussnlscha(\bm{\unlscha},\bm{\unlscha}') = \sqrt{\det\left(\frac{\Ybmnlscha}{2\pi}\right)}
    \exp\left\{-\frac{1}{4}\sum_{ab=1}^2 \unlscha_a \Tnlschaold{}_{ab}\unlscha_b \right.\\
    &\left.
    -\frac{1}{4}\sum_{ab=1}^2 \unlscha'_a \Tnlschaold{}_{ab}\unlscha'_b 
    + \sum_{ab=1}^2 \unlscha_a \Anlschaold{}_{ab}\unlscha'_b \right\}
\end{aligned}
\end{equation}
The NLSCHA tensors are related by $\Ybmnlscha = \Tbmnlscha - 2 \Abmnlscha$ and are defined by
\begin{subequations}
\label{eq app: Y def}
\begin{align}
    \Ynlscha{a}{b} & = \sum_{ij=1}^{2} \sqrt{\mathcal{M}}_{ai} 
    \overline{\Upsilon}_\text{nl,ij} \sqrt{\mathcal{M}}^T_{jb} \\
    \overline{\Upsilon}_\text{nl,ij} & = \sum_{\mu=1}^{2} \frac{2\onlschaold{}_\mu}{\hbar(1 + 2 \nnlscha{\mu})} \polnlscha{\mu}{i} \polnlscha{\mu}{j} 
\end{align}
\end{subequations}
and
\begin{subequations}
\label{eq app: A def}
\begin{align}
    \Anlschaold{}_{,ab} & = 
    \sum_{ij=1}^{2} \sqrt{\mathcal{M}}_{ai} \overline{A}_\text{nl,ij} \sqrt{\mathcal{M}}^T_{jb}  \\
    \overline{A}_\text{nl,ij} & = \sum_{\mu=1}^{2} \frac{2\onlschaold{}_\mu \nnlscha{\mu} ( 1+ \nnlscha{\mu})}{\hbar(1 + 2 \nnlscha{\mu})} \polnlscha{\mu}{i} \polnlscha{\mu}{j} 
\end{align}
\end{subequations}
where $\nnlscha{\mu}$ is the Bose-Einstein occupation number
\begin{equation}
\label{eq app: def n BE nlscha}
    \nnlscha{\mu} = \frac{1}{e^{\beta \hbar \onlscha{\mu}} -1}
\end{equation}
The square root of the mass tensor $\masstnsbm$ satisfies
\begin{equation}
\label{eq app: sqrt mass tns}
    \masstns{ab} = 
    \sum_{i=1}^{2} \sqrtmasstns{ai} \sqrtmasstnsT{ib}
\end{equation}

It seems that there is a contradiction within nonlinear SCHA because $\gaussnlscha(\bm{\unlscha}, \bm{\unlscha}')$ is normalized as a Gaussian \cite{INPREPARAZIONE} but the radial variable $\unlscha_1$ is defined between $[0,+\infty)$ (see Eq.\ \eqref{def: stereographic projection in 2D}). So we approximate the normalization condition extending the range of $\unlscha_1$ in $(-\infty,+\infty)$
\begin{equation}
\label{eq app: normalization nlscha}
\begin{aligned}
    \Tr\left[\rhohatnlscha\right] 
    & = \int_{-\infty}^{+\infty}\hspace{-0.2cm} d\bm{R} \bra{\bm{R}} \rhohatnlscha \ket{\bm{R}} \\
    & \simeq \int_{-\infty}^{+\infty}\hspace{-0.2cm} d\bm{\unlscha} \gaussnlscha(\bm{\unlscha}, \bm{\unlscha}) = 1
\end{aligned}
\end{equation}
The above assumption is justified by the following argument. In a diatomic molecule, the linear vibration along the radial coordinate $\unlscha_1$ is a high energy mode so that the bond length is always well defined hence the corresponding component of the wave function is very localized, i.e.\ it decays rapidly to zero, otherwise the atoms of the molecule will collapse one into the other. No approximations are considered for the normalization along $\unlscha_2$ as we used the stereographic projection that maps the angles $[0,2\pi]$ in $(-\infty,+\infty)$ (see Eq.\ \eqref{eq: phi - phi0 nlscha}).

\section{The nonlinear SCHA free energy}
\label{APP: The nonlinear SCHA free energy}
In appendix \ref{APP SUBSEC: Preliminary definitions} we compute all the necessary quantities to get the NLSCHA kinetic energy according to Ref.\ \cite{INPREPARAZIONE}. In appendix \ref{APP SUBSEC: Free energy calculation} we present the full NLSCHA free energy following Ref.\ \cite{INPREPARAZIONE}.

\subsection{Preliminary definitions}
\label{APP SUBSEC: Preliminary definitions}
We define the Jacobian of $\xibmnlscha$ (Eq.\ \eqref{def: stereographic projection in 2D})
\begin{equation}
\label{eq app: def J dR du}
    \Jnlscha{i}{j} = \pdv{R_i}{\unlscha_j}
    = \begin{bmatrix}
        \pdv{x}{\unlscha_1} & \pdv{y}{\unlscha_1} \\
        \pdv{x}{\unlscha_2} & \pdv{y}{\unlscha_2} 
    \end{bmatrix}
    = 
    \begin{bmatrix}
        \frac{x - \xcnlscha}{\unlscha_1 + \rOnlscha} & \frac{y - \ycnlscha}{\unlscha_1 + \rOnlscha} \\
        -\frac{y - \ycnlscha}{\rOnlscha \fsquarenlscha} & -\frac{x - \xcnlscha}{\rOnlscha \fsquarenlscha} 
    \end{bmatrix}
\end{equation}
where $\fnlscha$ is
\begin{equation}
\label{eq app: def f function nlscha}
     \fnlscha=\sqrt{1+\left(\frac{\unlscha_2}{2 \rOnlscha}\right)^2}
\end{equation}
The determinat of Eq.\ \eqref{eq app: def J dR du}
\begin{equation}
    \detJnlscha =
    \left|\det\left(\Jbmnlscha\right)\right|
    =\frac{\unlscha_1 + \rOnlscha}{\rOnlscha\fsquarenlscha}
\end{equation}

The inverse metric tensor $\gbmnlscha$ is
\begin{equation}
\label{eq app: def g}
    \gnlscha{a}{b} = \sum_{i=1}^2 \pdv{\unlscha_a}{R_i } \pdv{\unlscha_b}{R_j} = 
    \begin{bmatrix}
    1 & 0 \\
    0 & (\rOnlscha \fnlscha)^2 \hfunnlscha
    \end{bmatrix}
\end{equation}
where $\hfunnlscha$ is
\begin{equation}
    \hfunnlscha = \left(\frac{\fnlscha}{\unlscha_1 + \rOnlscha}\right)^2
\end{equation}
In addition, we define the vector $\dlogJdqbm$ as
\begin{equation}
\label{eq app: def d log d u}
    \dlogJdqbm = \frac{1}{2}\pdv{\log(\detJnlscha)}{\bm{\unlscha}} = \begin{bmatrix}
    \frac{1}{2(\unlscha_1 + \rOnlscha)} \\ 
    -\frac{\unlscha_2}{(2 \rOnlscha \fnlscha)^2 }
    \end{bmatrix}
\end{equation}

\subsection{Free energy calculation}
\label{APP SUBSEC: Free energy calculation}
According to Ref.\ \cite{INPREPARAZIONE} the nonlinear SCHA kinetic energy is
\begin{equation} 
\label{eq app: kinetic final}
    \hspace{-0.2cm}
    \Tr\left[\frac{\hat{\bm{P}}^2}{2m} \rhohatnlscha\right] =\left\langle
    \Knlscha \right\rangle_\text{nl}
    = \myint d\bm{\unlscha} \Knlscha
     \gaussnlscha(\bm{\unlscha})
\end{equation}
$\Knlscha$ is the kinetic kernel \cite{INPREPARAZIONE}
\begin{equation}
\label{eq app: knlscha}
    \Knlscha = \sum_{a=1}^{2} \left(\Ktwonlscha{a}
    \Ltwonlscha{a}  + \Konenlscha{a}\Lonenlscha{a} \right)+\Kzeronlscha
\end{equation}
where $\Lonebmnlscha$, $\Ltwobmnlscha$ are defined by
\begin{subequations}
\label{eq app: def L2 L1 coeff}
\begin{align}
    \Lonenlscha{a} &=  - \pdv{ \log\left(\gaussnlscha(\bm{\unlscha})\right)}{\widetilde{\unlscha}_a} = 
       \frac{1}{\sqrt{m}}\sum_{i=1}^{3N} \Ynlscha{a}{i} \unlscha_i  \label{eq app: L one app}   \\
    \Ltwonlscha{a} & = 
    -  \pdv{ \log\left(\gaussnlscha(\bm{\unlscha})\right)}{\widetilde{\unlscha}_a}{\widetilde{\unlscha}_a} 
    \label{eq app: L two def}\\
    & = 
    \frac{1}{m}\left(\Ynlscha{a}{a}- \sum_{ij=1}^{3N}
    \Ynlscha{a}{i} \Ynlscha{a}{j}
    \unlscha_i \unlscha_j\right)
    \notag
\end{align}
\end{subequations}
and $\Kzeronlscha$, $\Konebmnlscha$, and $\Ktwobmnlscha$  by
\begin{subequations}
\label{eq app: K2 K1 K0 def}
\begin{align}
    \Kzeronlscha  & = \frac{\hbar^2}{2m}\left\{
    \Tr\left[\gbmnlscha \cdot \left(\frac{\Ybmnlscha}{4}
    + \Abmnlscha\right)\right] +\dlogJdqbm \cdot \gbmnlscha \cdot \dlogJdqbm \right\} \notag\\
     & = \frac{\hbar^2}{2m}
    \Tr\left[\gbmnlscha \cdot \left(\frac{\Ybmnlscha}{4}
    + \Abmnlscha\right)\right]
    + \frac{\hbar^2\hfunnlscha}{8m}  \\
    \Konebmnlscha & = \frac{\hbar^2}{4} \gbmnlscha \cdot \dlogJdqbm
    = \frac{\hbar^2}{4\sqrt{m}}
    \begin{bmatrix}
       \frac{1}{2(\unlscha_1 + \rOnlscha)}\\  
       -\frac{1}{4}  \unlscha_2 \hfunnlscha
    \end{bmatrix} \\ 
    \Ktwobmnlscha
     & = -\frac{\hbar^2}{8} \text{diag}(\gbmnlscha)
    = -\frac{\hbar^2}{8}
     \begin{bmatrix}
    1\\ (\rOnlscha \fnlscha)^2 \hfunnlscha
    \end{bmatrix}
\end{align}
\end{subequations}
where $\gbmnlscha$ and $\dlogJdqbm$ are defined in Eqs \eqref{eq app: def g} \eqref{eq app: def d log d u}. The potential energy is
\begin{equation}
\begin{aligned}
    \Tr\left[\hat{V}^\text{(BO)} \rhohatnlscha\right]
    & = \int_{-\infty}^{+\infty} d\bm{\unlscha} V^\text{(BO)}(\xibmnlscha(\bm{\unlscha})) \gaussnlscha(\bm{\unlscha}) \\
    & = \averagegaussnl{V^\text{(BO)}}
\end{aligned}
\end{equation}
So the NLSCHA free energy is
\begin{equation}
\label{eq app: nlscha free energy}
    \Fnl = \left\langle
    \Knlscha \right\rangle_\text{nl} +\averagegaussnl{V^\text{(BO)}} - TS_\text{nl}
\end{equation}
where the entropy is harmonic as discussed in Ref.\ \cite{INPREPARAZIONE}
\begin{equation}
\begin{aligned}
    & S_\text{nl} = \Tr\left[\rhohatnlscha \log(\rhohatnlscha)\right] \\
    & = k_\text{B}\sum_{\mu=1}^2\left[(1+\nnlscha{\mu})\log(1+\nnlscha{\mu}) - \nnlscha{\mu}\log(\nnlscha{\mu})\right]
\end{aligned}
\end{equation}

\section{Gradient of the nonlinear SCHA free energy}
\label{APPENDIX: Nonlinear SCHA gradient}
We start with the gradient with respect to the center of the curvature $\RCnlscha$ (Eq.\ \eqref{eq: R_C})
\begin{equation}
    \pdv{\Fnl}{\RCnlscha} = \averagegaussnl{\pdv{V^{\text{(BO)}}}{\bm{R}}}
\end{equation}
The gradient with respect to curvature vector $\RTnlscha$ (Eq.\ \eqref{eq: R_T}) is the following
\begin{equation}
\begin{aligned}
    & \pdv{\Fnl}{\RTnlscha} = \averagegaussnl{\pdv{\Knlscha}{\RTnlscha}}+ 
    \averagegaussnl{\pdv{V^{\text{(BO)}}}{\RTnlscha}} \\
    & = \frac{\RTnlscha}{|\RTnlscha|} \averagegaussnl{\pdv{\Knlscha}{\rOnlscha}} 
    + \sum_{i=1}^2 \pdv{R_i}{\RTnlscha} \averagegaussnl{\pdv{V^{\text{(BO)}}}{{R_i}}}
\end{aligned}
\end{equation}
noting that $\Knlscha(\bm{\unlscha})$ (Eq.\ \eqref{eq app: knlscha}) depends only on the curvature $|\RTnlscha|=\rOnlscha=\kappa^{-1}$. The derivative of the kinetic energy kernel $\Knlscha$ (Eq.\ \eqref{eq app: knlscha}) is
\begin{equation}
\begin{aligned}
    &\pdv{\Knlscha}{\rOnlscha} = 
    \sum_{a=1}^{2} \left(
    \pdv{\Ktwonlscha{a}}{\rOnlscha} \Ltwonlscha{a}  + \pdv{\Konenlscha{a}}{\rOnlscha} \Lonenlscha{a}\right)  
    + \pdv{\Kzeronlschaold}{\rOnlscha} (\bm{\unlscha})
\end{aligned}
\end{equation}
where $\rOnlscha$ the derivative of the coefficients (Eqs \eqref{eq app: K2 K1 K0 def}) are
\begin{subequations}
\begin{align}
    \pdv{\Kzeronlscha}{\rOnlscha} =  & 
    \frac{\hbar^2}{2m}
    \Tr\left[\pdv{\gbmnlscha}{\rOnlscha} \cdot \left(\frac{\Ybmnlscha}{4}
    + \Abmnlscha\right)\right] \notag\\
    &    + \frac{\hbar^2}{8m} \pdv{\hfunnlscha}{\rOnlscha}
    \\
    \pdv{\Konebmnlscha}{\rOnlscha} = &   
    \frac{\hbar^2}{4\sqrt{m}} \begin{bmatrix}
    \pdv{}{\rOnlscha} \frac{1}{2(\unlscha_1 + \rOnlscha)} \\
    -\frac{1}{4} \unlscha_2 \pdv{}{\rOnlscha} \hfunnlscha 
    \end{bmatrix}\\
    \pdv{\Ktwobmnlscha}{\rOnlscha} =  &
    -\frac{\hbar^2}{8}\begin{bmatrix}
    0 \\
    \pdv{[(r_0 \fnlscha)^2 \hfunnlscha]}{\rOnlscha}
    \end{bmatrix}\\
    \pdv{\gbmnlscha}{\rOnlscha} = & 
    \begin{bmatrix}
    0 & 0 \\
    0 & \pdv{[(\rOnlscha \fnlscha)^2 \hfunnlscha]}{\rOnlscha}
    \end{bmatrix} \\
    \pdv{\fsquarenlscha}{\rOnlscha} = & -\frac{\unlscha_2{}^2}{2 \rOnlscha^3 }\\
    \pdv{\hfunnlscha}{\rOnlscha} = & \frac{1}{(\unlscha_1+\rOnlscha)^2} \pdv{\fsquarenlscha}{\rOnlscha}
    - \frac{2\hfunnlscha}{\rOnlscha + \unlscha_1}
\end{align}
\end{subequations}
The gradient of the Cartesian coordinates $\bm{R}$ with respect to the curvature vector $\RTnlscha$ is
\begin{equation}
    \pdv{R_i}{\RTnlscha} = \frac{\RTnlscha}{|\RTnlscha|}\pdv{R_i}{\rOnlscha}  
    + \pdv{\phiOnlscha}{\RTnlscha} \pdv{R_i}{\phiOnlscha} 
\end{equation}
where the derivatives of the Cartesian coordinates $\bm{R} = (x, y)$ are
\begin{equation}
    \begin{bmatrix}
        \pdv{x}{\rOnlscha} & \pdv{y}{\rOnlscha} \\
        \pdv{x}{\phiOnlscha} & \pdv{y}{\phiOnlscha} 
    \end{bmatrix}
    =
    \begin{bmatrix}
        \frac{x - \xcnlscha}{\unlscha_1 + \rOnlscha} + \frac{\unlscha_2(y -\ycnlscha)}{\rOnlscha^2 \fsquarenlscha} &
        \frac{y - \ycnlscha}{\unlscha_1 + \rOnlscha} - \frac{\unlscha_2(x - \xcnlscha)}{\rOnlscha^2 \fsquarenlscha} \\
        -(y - \ycnlscha) & +(x - \xcnlscha)
    \end{bmatrix}
\end{equation}
and the derivative of $\phiOnlscha$ is computed from the definition of $\RTnlscha$ (Eq.\ \eqref{eq: R_T})
\begin{equation}
    \pdv{\phiOnlscha}{\RTnlscha} = -\frac{1}{|\RTnlscha|^2}
    \begin{bmatrix}
        \RTcmpnlscha{2} & -\RTcmpnlscha{1}
    \end{bmatrix}
\end{equation}

To compute the gradient with respect to $\FCbmnlscha$ we use the following formula introduced by Ref.\ \cite{Bianco}
\begin{equation}
\label{eq app: derivative wrt FC nlscha}
\begin{aligned}
    & \pdv{\averagegaussnl{O}}{\FCbmnlscha} =  
    \averagegaussnl{\pdv{O}{\FCbmnlscha}} \\
    & +
    \frac{1}{2} \sum_{ijk=1}^2 
    \pdv{\Yinvnlscha{i}{j}}{\FCbmnlscha} 
    \Ynlscha{i}{k} \averagegaussnl{\unlscha_k \pdv{O}{\unlscha_j}} 
\end{aligned}
\end{equation}
The first term of Eq.\ \eqref{eq app: derivative wrt FC nlscha} takes into account the explicit dependence of $O$ on $\FCbmnlscha$ while the second considers the change in the probability distribution $\gaussnlscha(\bm{\unlscha})$. With Eq.\ \eqref{eq app: derivative wrt FC nlscha}, the derivative of $\Fnl$ with respect to the auxiliary force constant matrix $\FCbmnlscha$ is
\begin{equation}
\label{eq app: d Fnl d FC}
\begin{aligned}
    & \pdv{\Fnl}{\FCbmnlscha} = 
    \averagegaussnl{\pdv{\Knlscha(\bm{\unlscha})}{\FCbmnlscha}}  
    \\
    &
    +\frac{1}{2} \sum_{ijk=1}^2 \pdv{\Yinvnlscha{i}{j}}{\FCbmnlscha} \Ynlscha{i}{k} \averagegaussnl{\unlscha_k \left( 
    \pdv{\Knlscha}{\unlscha_j} 
    + \pdv{V^{\text{(BO)}}}{\unlscha_j} \right)}  
    \\
    &
    - T \pdv{S_\text{nl}}{\FCbmnlscha}
\end{aligned}
\end{equation}
where derivative of the entropy $S_\text{nl}$ with respect to $\FCbmnlscha$ is computed in Ref.\ \cite{INPREPARAZIONE}.

The first term is the derivative of $\Knlscha(\bm{\unlscha})$ (Eq.\ \eqref{eq app: knlscha}) with respect to $\FCbmnlscha$
\begin{equation}
    \pdv{\Knlscha}{\FCbmnlscha} =
    \sum_{a=1}^{2} \left(
    \Ktwonlscha{a} \pdv{\Ltwonlscha{a}}{\FCbmnlscha}
    + \Konenlscha{a}\pdv{\Lonenlscha{a}}{\FCbmnlscha} \right) + \pdv{\Kzeronlscha}{\FCbmnlscha}
\end{equation}
The derivatives of $\Lonebmnlscha$, $\Ltwobmnlscha$ (Eqs \eqref{eq app: def L2 L1 coeff}) are
\begin{subequations}
\begin{align}
    \pdv{\Lonenlscha{a}}{\FCbmnlscha}   & = \frac{1}{\sqrt{m}}\left(
    \sum_{i=1}^{2} \pdv{\Ynlscha{a}{i}}{\FCbmnlscha} \unlscha{}_i  \right) \\
    \pdv{\Ltwonlscha{a}}{\FCbmnlscha}   & = \frac{1}{m}
    \left(\pdv{\Ynlscha{a}{a}}{\FCbmnlscha}  - 2\sum_{ij=1}^{2}
    \pdv{\Ynlscha{a}{i}}{\FCbmnlscha} \Ynlscha{a}{j} \unlscha{}_i \unlscha{}_j  \right)
\end{align}
\end{subequations}
The derivative of $\Kzeronlscha$ (Eq.\ \eqref{eq app: K2 K1 K0 def}) is
\begin{equation}
    \pdv{\Kzeronlscha}{\FCbmnlscha} = 
     \frac{\hbar^2}{2m}
    \Tr\left[\gbmnlscha \cdot \left(\frac{1}{4}
    \pdv{\Ybmnlscha}{\FCbmnlscha}
    + \pdv{\Abmnlscha}{\FCbmnlscha}\right)\right]
\end{equation}
The derivatives of $\Ybmnlscha$/$\Abmnlscha$ can be computed with the expressions of Ref.\ \cite{INPREPARAZIONE}.

The derivative of the kinetic energy kernel $\Knlscha(\bm{\unlscha})$ with respect to the auxiliary displacements $\bm{\unlscha}$ is
\begin{equation}
\begin{aligned}
    \pdv{\Knlscha}{\bm{\unlscha}} = & 
    \sum_{a=1}^{2}
    \left(\pdv{\Ktwonlscha{a}}{\bm{\unlscha}} \Ltwonlscha{a}
    +\Ktwonlscha{a} \pdv{\Ltwonlscha{a} }{\bm{\unlscha}}\right) \\
    & +  \sum_{a=1}^{2}
    \left(\pdv{\Konenlscha{a}}{\bm{\unlscha}} \Lonenlscha{a}
    +\Konenlscha{a} \pdv{\Lonenlscha{a} }{\bm{\unlscha}}\right)  + \pdv{\Kzeronlscha}{\bm{\unlscha}} 
\end{aligned}
\end{equation}  
The derivatives with respect to $\bm{\unlscha}$ of  $\Kzeronlscha$, $\Konebmnlscha$, and $\Ktwobmnlscha$  (Eqs \eqref{eq app: K2 K1 K0 def}) are
\begin{subequations}
\begin{align}
    \pdv{\Kzeronlscha}{\bm{\unlscha}} = &  
    \frac{\hbar^2}{2m}\left\{
    \Tr\left[\pdv{\gbmnlscha}{\bm{\unlscha}} \cdot \left(\frac{\Ybmnlscha}{4}
    + \Abmnlscha\right)\right] \right. \notag\\
    &\left.
    + \frac{1}{4} \pdv{\hfunnlscha}{\bm{\unlscha}}\right\}
     \\
      \pdv{\Konebmnlscha}{\bm{\unlscha}} = &   
    \frac{\hbar^2}{4\sqrt{m}} \begin{bmatrix}
       \pdv{}{\bm{\unlscha}} \frac{1}{2(\unlscha_1 + \rOnlscha)} \\
       -\frac{1}{4} \unlscha_2 \pdv{}{\bm{\unlscha}} \hfunnlscha 
    \end{bmatrix}\\
     \pdv{\Ktwobmnlscha}{\bm{\unlscha}} = & 
    -\frac{\hbar^2}{8}\begin{bmatrix}
    0 \\
    \rOnlscha^2 \pdv{(\fsquarenlscha \hfunnlscha)}{\bm{\unlscha}}
    \end{bmatrix}\\
    \pdv{\gbmnlscha}{\bm{\unlscha}} = &
    \begin{bmatrix}
    0 & 0 \\
    0 & \rOnlscha^2 \pdv{[\fsquarenlscha \hfunnlscha]}{\bm{\unlscha}}
    \end{bmatrix}
     \\
     \pdv{\fsquarenlscha}{\bm{\unlscha}} = &
     \begin{bmatrix}
         0 \\
         \frac{\unlscha_2}{2\rOnlscha^2}
     \end{bmatrix}\\
     \pdv{\hfunnlscha}{\bm{\unlscha}} = &
    \begin{bmatrix}
        -\frac{2}{(\rOnlscha + \unlscha_1)} \hfunnlscha \\
        \frac{1}{(\rOnlscha + \unlscha_1)^2} \pdv{\fsquarenlscha}{\unlscha_2}
    \end{bmatrix}
\end{align}
\end{subequations}
and the derivatives of $\Lonebmnlscha$, $\Ltwobmnlscha$ (Eqs \eqref{eq app: def L2 L1 coeff}) are
\begin{subequations}
\label{eq app: diff L 2 1 wrt u}
\begin{align}
    \pdv{\Lonenlscha{a}}{\unlscha_b}  = & \frac{1}{\sqrt{m}}
   \Ynlscha{a}{b}\\
    \pdv{\Ltwonlscha{a}}{\unlscha_b}  =  &
   \frac{1}{m}\left( - 2 \Ynlscha{a}{b} \sum_{i=1}^2\Ynlscha{a}{i} \unlscha_i  \right) 
\end{align}
\end{subequations}
The derivative of the BOES with respect to the auxiliary displacements $\bm{\unlscha}$ is
\begin{equation}
    \pdv{V^{\text{(BO)}}(\bm{R})}{\unlscha_j} = - \sum_{i=1}^2
    \Jnlscha{i}{j} f^\text{(BO)}_i
\end{equation}
where $\Jbmnlscha$ is defined in Eq.\ \eqref{eq app: def J dR du}. 

We minimize with respect to $\sqrtmasstnsbm$ so that we can enforce the positive definiteness of $\masstnsbm = \sqrtmasstnsTbm \cdot \sqrtmasstnsbm$
\begin{equation}
\label{eq app: d Fnl d sqrtmass}
\begin{aligned}
    & \pdv{\Fnl}{\sqrtmasstnsbm} =  \averagegaussnl{\pdv{\Knlscha}{\sqrtmasstnsbm}}  
    \\
    &+\frac{1}{2} \sum_{ijk=1}^2
    \pdv{\Yinvnlscha{i}{j}}{\sqrtmasstnsbm} 
    \Ynlscha{i}{k} \averagegaussnl{\unlscha_k \left(\pdv{\Knlscha}{\unlscha_j} + \pdv{V^{\text{(BO)}}}{\unlscha_j} \right)}   \\
    &
    - T \pdv{S_\text{nl}}{\sqrtmasstnsbm}
\end{aligned}
\end{equation}
This equation is a straightforward generalization of Eq.\ \eqref{eq app: d Fnl d FC}. The derivative $\Yinvbmnlscha$ is computed in Ref.\ \cite{INPREPARAZIONE}. The derivative of the kinetic kernel is
\begin{equation}
    \pdv{\Knlscha}{\sqrtmasstnsbm} =
    \sum_{a=1}^{2} \left(
    \Ktwonlscha{a} \pdv{\Ltwonlscha{a}}{\sqrtmasstnsbm}
    + \Konenlscha{a}\pdv{\Lonenlscha{a}}{\sqrtmasstnsbm} \right) + \pdv{\Kzeronlscha}{\sqrtmasstnsbm}
\end{equation}
Then, to compute
\begin{subequations}
\begin{align}
    \pdv{\Ltwonlscha{a}}{\sqrtmasstnsbm}   & = 
    \pdv{\Ynlscha{a}{a}}{\sqrtmasstnsbm}  - 2\sum_{ij=1}^{2}
    \pdv{\Ynlscha{a}{i}}{\sqrtmasstnsbm} \Ynlscha{a}{j} \unlscha{}_i \unlscha{}_j \\
     \pdv{\Lonenlscha{a}}{\sqrtmasstnsbm}   & = 
    \sum_{i=1}^{2} \pdv{\Ynlscha{a}{i}}{\sqrtmasstnsbm} \unlscha{}_i  
\end{align}
\end{subequations}
and
\begin{equation}
    \pdv{\Kzeronlscha}{\sqrtmasstnsbm} =  
    \frac{\hbar^2}{2m}
    \Tr\left[\gbmnlscha \cdot \left(
    \frac{1}{4} \pdv{\Ybmnlscha}{\sqrtmasstnsbm}
    + \pdv{\Abmnlscha}{\sqrtmasstnsbm}\right)\right]
\end{equation}
we employ the formulas derived in Ref.\ \cite{INPREPARAZIONE}. The derivative of the entropy is
\begin{equation}
\begin{aligned}
    & \pdv{S_\text{nl}}{\sqrt{\mathcal{M}}_{ab}} = \sum_{ij=1}^{2}
    \pdv{S_\text{nl}}{\Dnlschaold{}_{,ij}} \pdv{\Dnlschaold{}_{,ij}}{\sqrt{\mathcal{M}}_{ab}}
\end{aligned}
\end{equation}
where $\pdv{\Dbmnlscha}{\sqrtmasstnsbm}$  and $\pdv{S_\text{nl}}{\Dbmnlscha}$ are given by Ref.\ \cite{INPREPARAZIONE}.

\section{Nonlinear SCHA simulations}
\label{APP: Nonlinear SCHA simulations}
The NLSCHA simulations were performed on a uniform square grid in $\bm{\unlscha}$ space of size $1000$ between $\pm 3$ Bohr. The conjugate gradient (CG) minimization of the NLSCHA free energy was performed with the \texttt{scipy} \cite{scipy} function \texttt{scipy.optimize.minimize} setting \texttt{gtol} to $10^{-8}$ and \texttt{maxiter} $4000$. We report the free parameters of NLSCHA both at zero (appendix \ref{APP: NLSCHA parameters zero temperature}) and finite temperature (appendix \ref{APP: NLSCHA parameters at finite temperature}). 

\subsection{Parameters at zero temperature}
\label{APP: NLSCHA parameters zero temperature}
\begin{figure}[!htb]
    \centering
    \begin{minipage}[c]{1.0\linewidth}
    \includegraphics[width=1.0\textwidth]{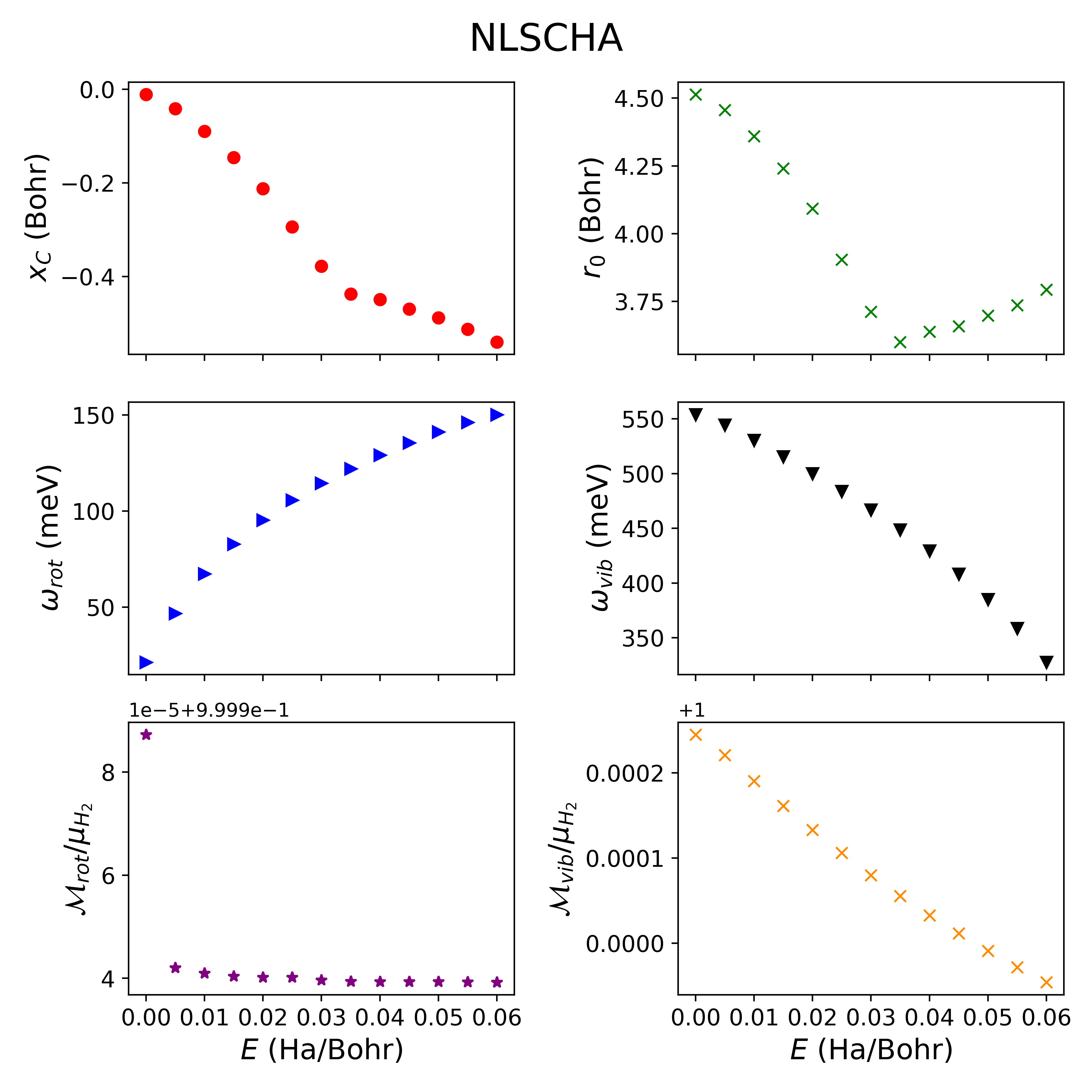}
    \end{minipage}
    \caption{The free parameters of NLSCHA ($X_C, r_0, \omega_\text{rot}, \omega_\text{vib}, \mathcal{M}_\text{rot}/\mu_{\ch{H2}}, \mathcal{M}_\text{vib}/\mu_{\ch{H2}}$) at $T=0$ K from $E=0$ Ha/Bohr to $E=0.06$ Ha/Bohr.}
    \label{fig:parameters nlscha zero T allE}
\end{figure}

\begin{figure}[!htb]
    \centering
    \begin{minipage}[c]{1.0\linewidth}
    \includegraphics[width=1.0\textwidth]{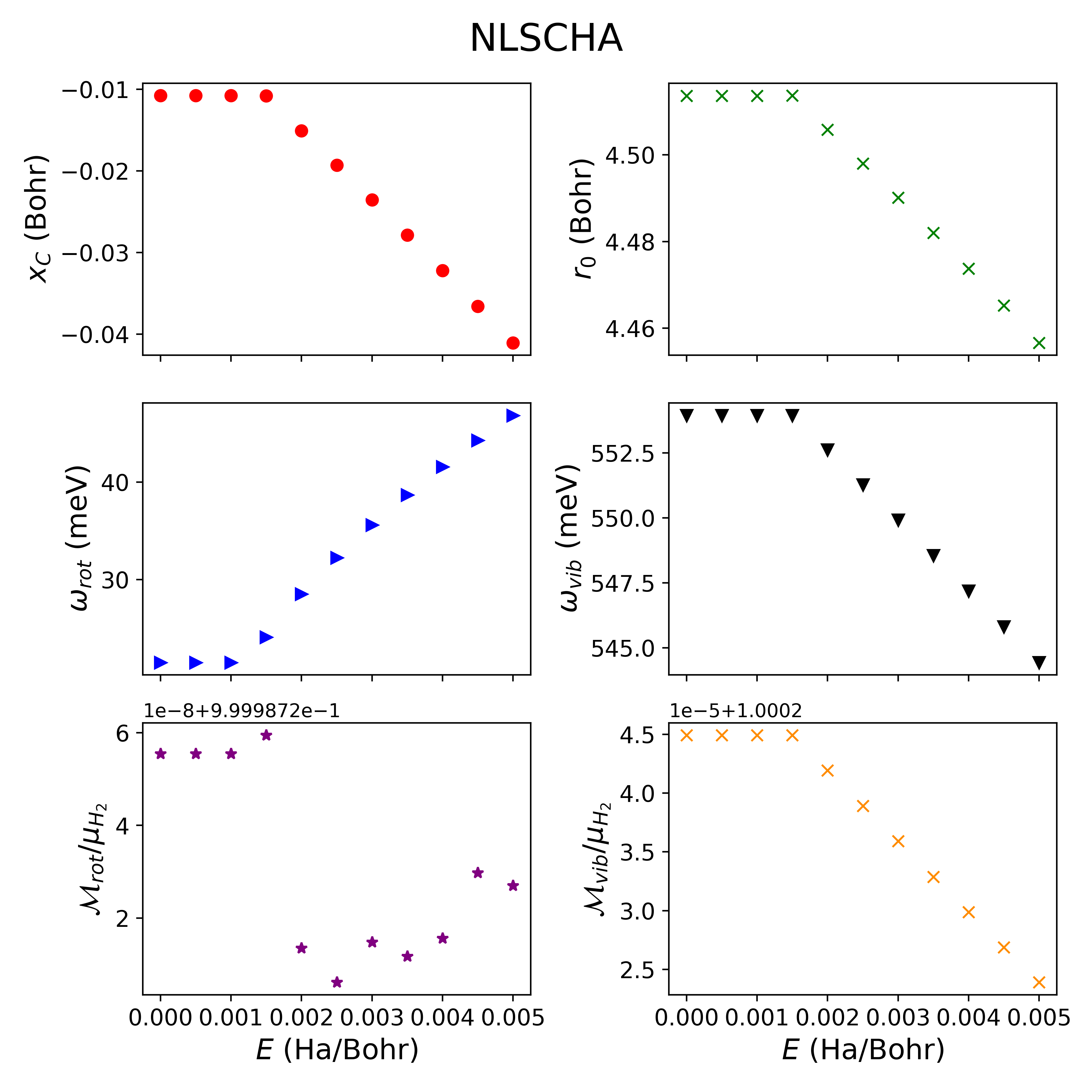}
    \end{minipage}
    \caption{The free parameters of NLSCHA ($X_C, r_0, \omega_\text{rot}, \omega_\text{vib}, \mathcal{M}_\text{rot}/\mu_{\ch{H2}}, \mathcal{M}_\text{vib}/\mu_{\ch{H2}}$) at $T=0$ K and low values of the crystal field $E$ from $E=0$ Ha/Bohr to $E=0.005$ Ha/Bohr.}
    \label{fig:parameters nlscha zero T allE zoom}
\end{figure}

In Figs \ref{fig:parameters nlscha zero T allE} \ref{fig:parameters nlscha zero T allE zoom} we report the nonlinear SCHA free parameters for $E=0-E=0.06$ Ha/Bohr and $E=0-E=0.005$ Ha/Bohr. We show the $x$-component of $\RCnlscha$, denoted by $\xcnlscha$, the radius of the curvature, $\rOnlscha$, the rotational and vibrational frequencies, $\omega_\text{rot}$ and $\omega_\text{vib}$, with the corresponding effective mass eigenvalues, $\mathcal{M}_\text{rot},\mathcal{M}_\text{vib}$, divided by the physical value $\mu_{\ch{H2}} = m_{\ch{H}}/2$. The symmetry of the potential, see Eq.\ \eqref{def: toy model potential}, fixes the values of $\ycnlscha=0$ Bohr and $\phiOnlscha=\pi$ are fixed. Note that at $T=0$ K we find that $\bm{\mathcal{M}}$ coincides with the physical one.

\newpage

\subsection{Parameters at finite temperature}
\label{APP: NLSCHA parameters at finite temperature}
\begin{figure}[!htb]
    \centering
    \begin{minipage}[c]{1.0\linewidth}
    \includegraphics[width=1.0\textwidth]{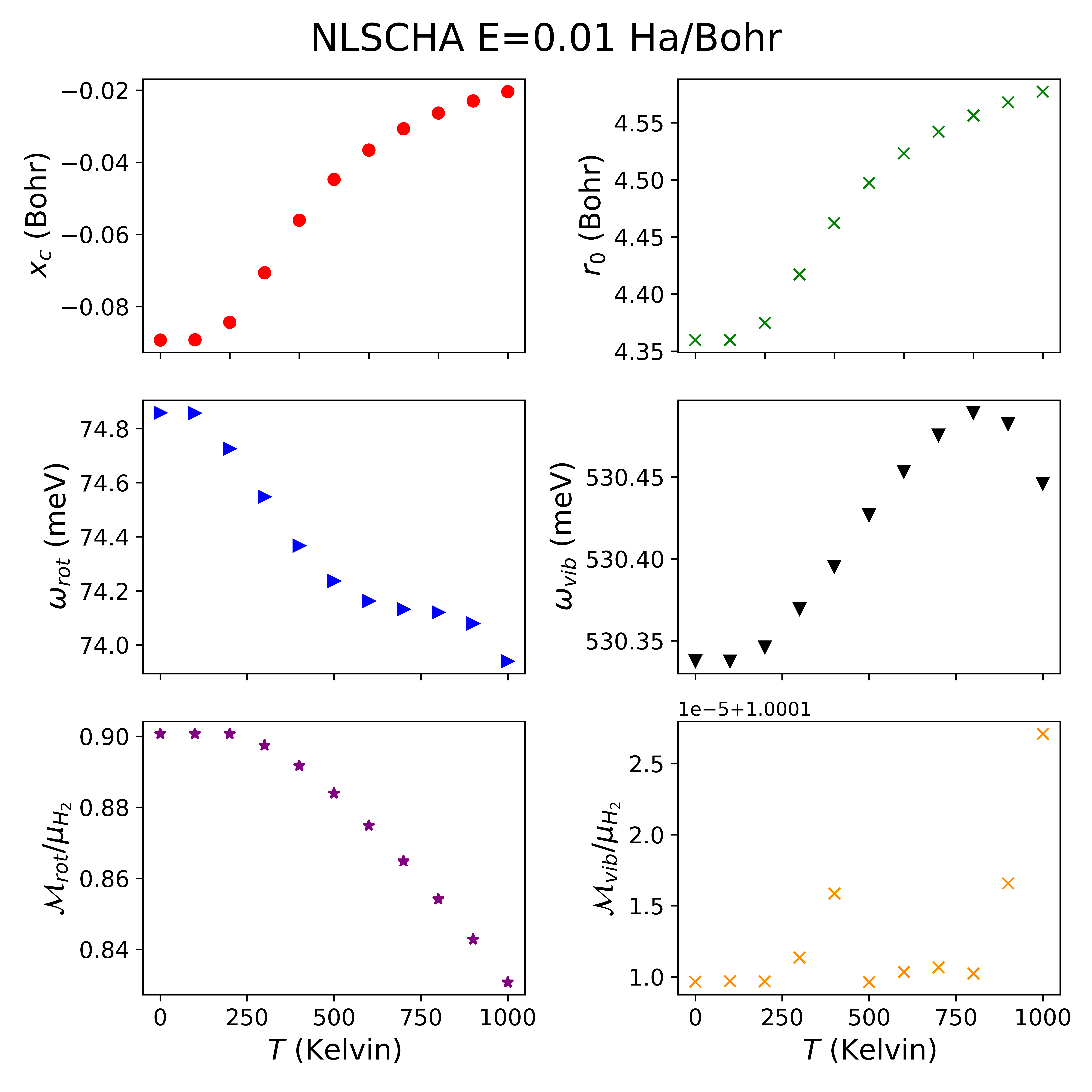}
    \end{minipage}
    \caption{The free parameters of NLSCHA ($X_C, r_0, \omega_\text{rot}, \omega_\text{vib}, \mathcal{M}_\text{rot}/\mu_{\ch{H2}}, \mathcal{M}_\text{vib}/\mu_{\ch{H2}}$) from $T=0$ K to $T=1000$ K with $E=0.01$ Ha/Bohr.}
    \label{fig:parameters nlscha finite T rot}
\end{figure}

\begin{figure}[!htb]
    \centering
    \begin{minipage}[c]{1.0\linewidth}
    \includegraphics[width=1.0\textwidth]{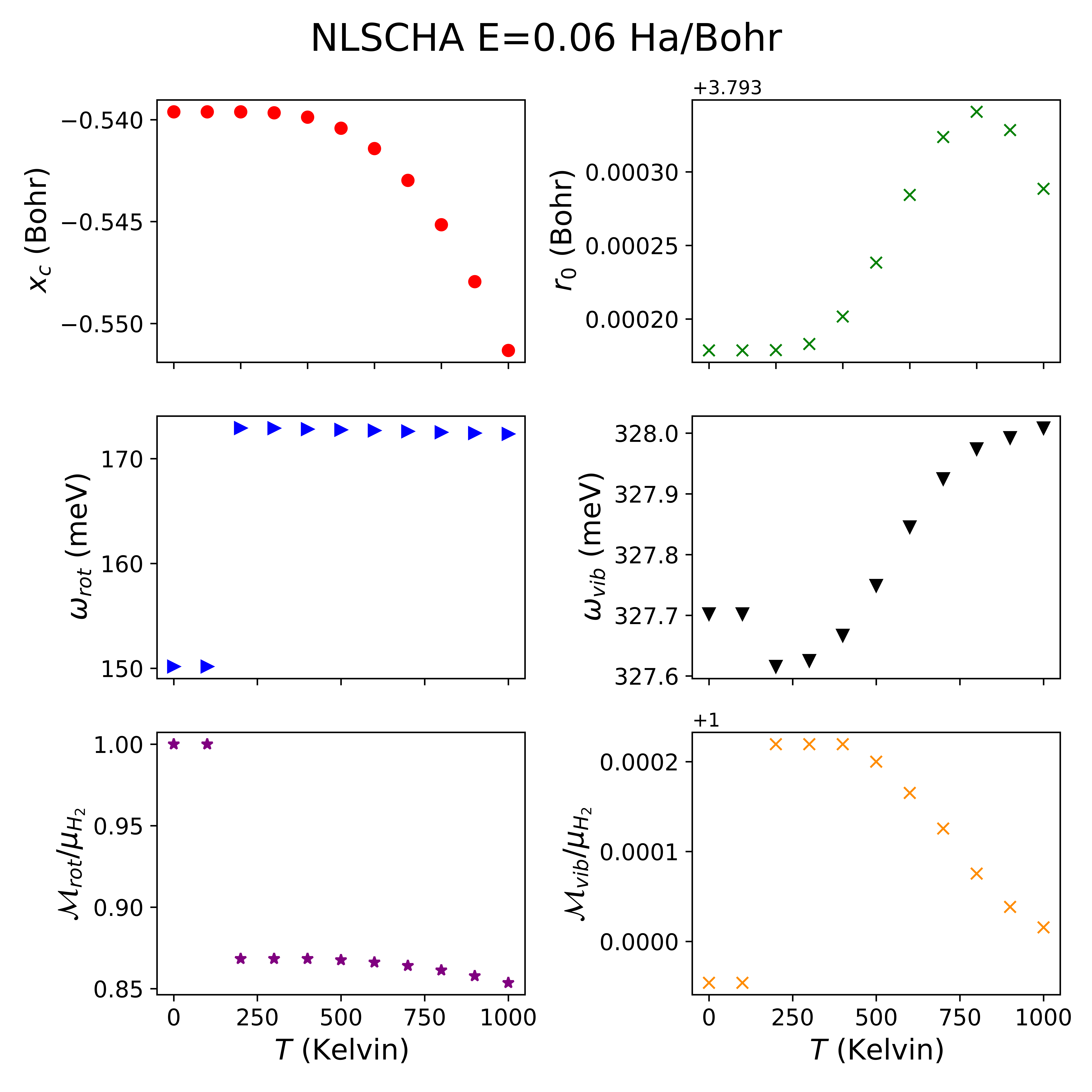}
    \end{minipage}
    \caption{The free parameters of NLSCHA ($X_C, r_0, \omega_\text{rot}, \omega_\text{vib}, \mathcal{M}_\text{rot}/\mu_{\ch{H2}}, \mathcal{M}_\text{vib}/\mu_{\ch{H2}}$) from $T=0$ K to $T=1000$ K with $E=0.06$ Ha/Bohr.}
    \label{fig:parameters nlscha finite T vib}
\end{figure}

Figs \ref{fig:parameters nlscha finite T rot} \ref{fig:parameters nlscha finite T vib} report the NLSCHA free parameters at finite temperature from $T=0$ K to $T=1000$ K for $E=0.01$ Ha/Bohr and $E=0.06$ Ha/Bohr. For $E=0.01$ Ha/Bohr, the eigenvalue of $\masstnsbm$ corresponding to the rotational mode, $\mathcal{M}_\text{rot}$, is lower than the physical one so that rotational fluctuations are enhanced. Note that for the vibron mode, the eigenvalue of $\masstnsbm$ is $\mathcal{M}_\text{vib} \simeq \mu_{\ch{H2}}$. For $E=0.06$ Ha/Bohr, $\mathcal{M}_\text{vib} < \mu_{\ch{H2}}$ at $200$ K compensates for the increase of $\omega_\text{rot}$.

\begin{figure}[!htb]
    \centering
    \begin{minipage}[c]{1.0\linewidth}
    \includegraphics[width=1.0\textwidth]{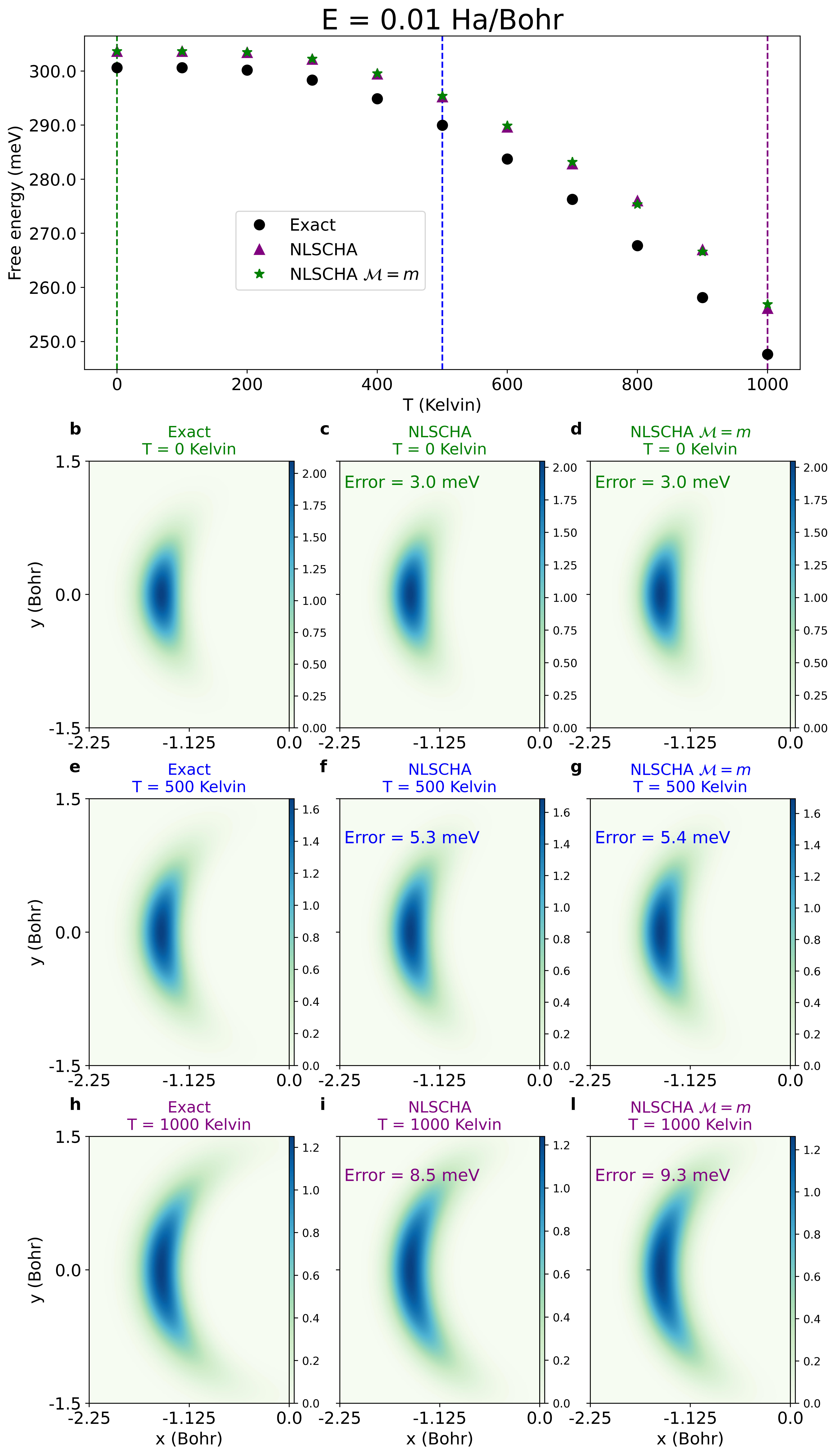}
    \end{minipage}
    \caption{Fig.\ a shows the exact, and NLSCHA free energies as a function of temperature for $E=0.01$ Ha/Bohr. We report the NLSCHA solutions with $\masstnsbm=\mu_\text{\ch{H2}}\bm{1}$ and $\masstnsbm\neq \mu_\text{\ch{H2}} \bm{1}$. Figs b-l report the exact (Figs b-e-h), and NLSCHA (Figs c-f-i with $\masstnsbm\neq \mu_\text{\ch{H2}}\bm{1}$ and Figs d-g-l with $\masstnsbm=\mu_\text{\ch{H2}}\bm{1}$) probability distribution $\rho(\bm{R})$ (Bohr$^{-2}$) at $0-500-1000$ K.}
    \label{fig: free energy nlscha mass tns analysis}
\end{figure}

In Fig.\ \ref{fig: free energy nlscha mass tns analysis} we compare the NLSCHA free energies ($E=0.01$ Ha/Bohr) obtained optimizing $\masstnsbm$ and keeping it fixed $\bm{\mathcal{M}} = \mu_{\ch{H2}} \bm{1}$. Fig.\ \ref{fig: free energy nlscha mass tns analysis} b-l report the corresponding probability distributions compared with the exact result. Note that at $T=1000$ K optimizing $\bm{\mathcal{M}}$ gives an error of $8.5$ meV while keeping it fixed yields $9.3$ meV.


\newpage
\section{Phase transition model}
\label{APP: Phase transition model}
\begin{figure*}[!htb]
    \centering
    \begin{minipage}[c]{0.8\linewidth}
    \includegraphics[width=1.0\textwidth]{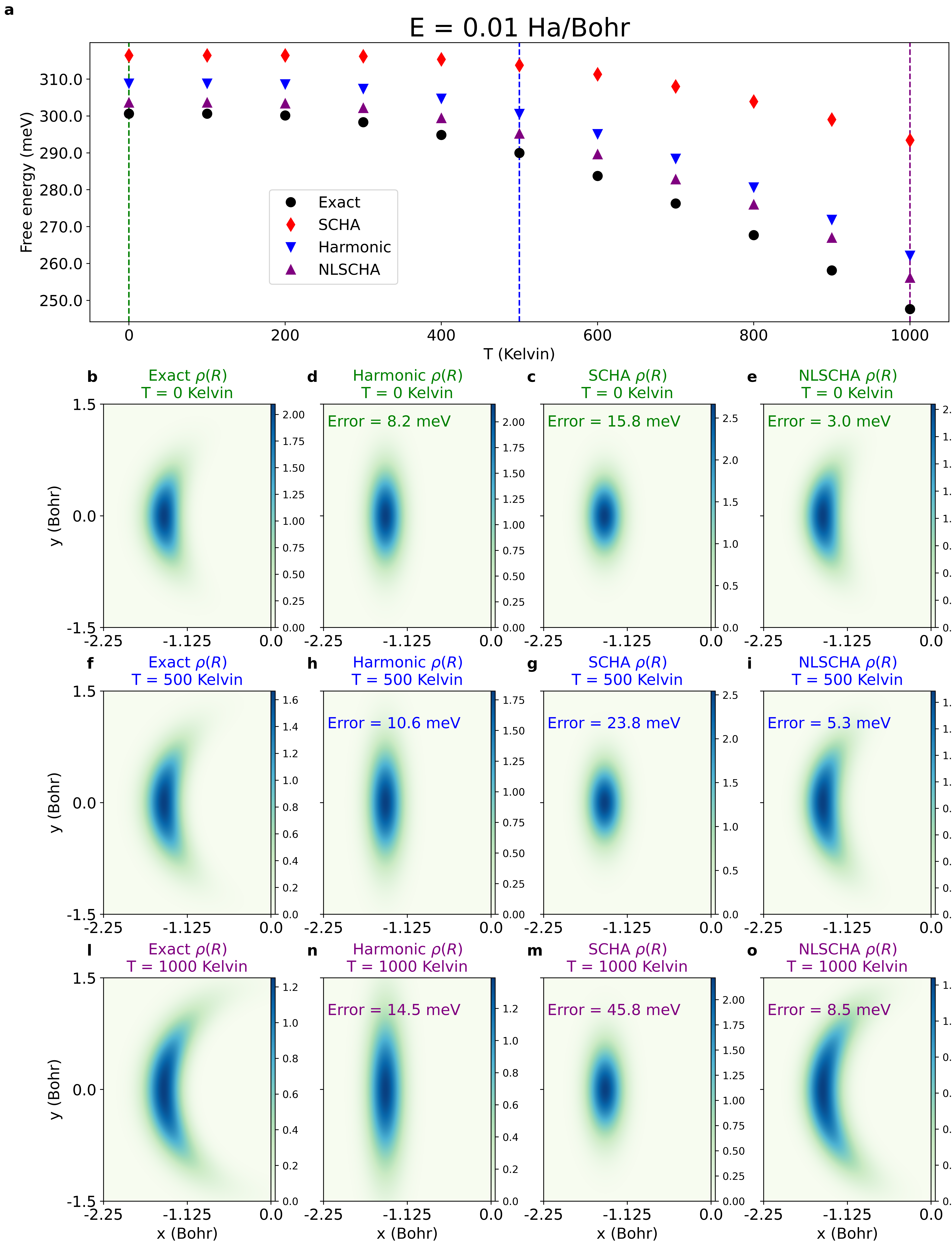}
    \end{minipage}
    \caption{Fig.\ a shows the exact, harmonic, SCHA and NLSCHA free energies as a function of temperature for $E=0.01$ Ha/Bohr. Figs b-o report the exact (Figs b-f-l), harmonic (Figs d-h-n), SCHA (Figs c-g-m) and NLSCHA (Figs e-i-o) probability distribution $\rho(\bm{R})$ (Bohr$^{-2}$) at $0-500-1000$ K.}
    \label{fig:free energy rotational}
\end{figure*}

\begin{figure*}[!htb]
    \centering
    \begin{minipage}[c]{0.8\linewidth}
    \includegraphics[width=1.0\textwidth]{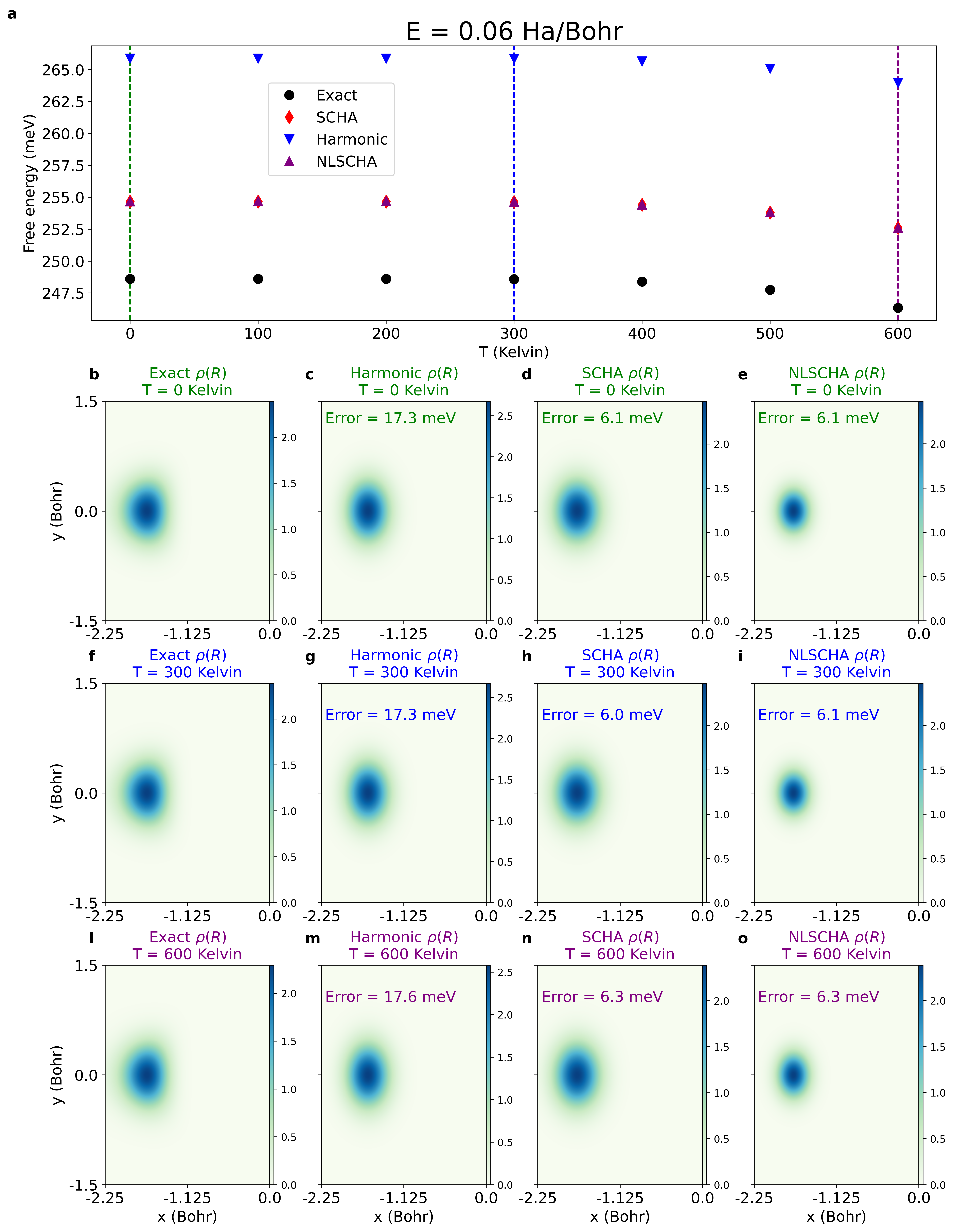}
    \end{minipage}
    \caption{Fig.\ a shows the exact, harmonic, SCHA and NLSCHA free energies as a function of temperature for $E=0.06$ Ha/Bohr. Figs b-o report the exact (Figs b-f-l), harmonic (Figs c-g-m), SCHA (Figs d-h-n) and NLSCHA (Figs e-i-o) probability distribution $\rho(\bm{R})$ (Bohr$^{-2}$) at $0-300-600$ K.}
    \label{fig:free energy vibrational}
\end{figure*}

In Fig.\ \ref{fig:free energy rotational}-\ref{fig:free energy vibrational} we report the exact, harmonic, SCHA and NLSCHA free energies for $E=0.01-0.06$ Ha/Bohr form $T=0$ K to $1000-600$ K along with the probability distributions. Note that when rotations are locked ($E=0.06$ Ha/Bohr) the NLSCHA solution coincides with the SCHA.

\clearpage
\newpage

\bibliography{apssamp}

\end{document}